\documentclass[11pt]{article}
\usepackage{amsmath,amssymb,color,graphics,epsfig}

\usepackage{amsfonts}

\newcommand{\be}{\begin{equation}}
\newcommand{\ee}{\end{equation}}
\newcommand{\bea}{\setlength\arraycolsep{2pt} \begin{eqnarray}}
\newcommand{\eea}{\end{eqnarray}}
\newcommand{\nn}{\nonumber}

\def\0{{\sst{(0)}}}
\def\1{{\sst{(1)}}}
\def\2{{\sst{(2)}}}
\def\3{{\sst{(3)}}}
\def\4{{\sst{(4)}}}
\def\5{{\sst{(5)}}}
\def\6{{\sst{(6)}}}
\def\7{{\sst{(7)}}}
\def\8{{\sst{(8)}}}
\def\sst#1{{\scriptscriptstyle #1}}

\begin{document}

\begin{flushright}
\end{flushright}

\vspace{25pt}
\begin{center}
{\large {\bf Exact fermionic Green's Function and Fermi surfaces from Holography \\ }}

\vspace{10pt}
Zhong-Ying Fan

\vspace{10pt}

{\it Department of Physics, Beijing Normal University, Beijing 100875, China}

\vspace{40pt}

\underline{ABSTRACT}
\end{center}

We construct a series of charged dilatonic black holes which share zero entropy in the zero temperature limit using Einstein-Maxwell-Dilaton theories.
In these black holes, the wave functions and the Green's functions of massless fermions can be solved exactly in terms of special functions in the phase space of $(\omega,k)$. We observe that for sufficiently large charge, there are many poles in the Green's function with vanishing $\omega$, which strongly signifies that Fermi surfaces exist in these holographic systems. The new distinguishing properties of the Green's function arising in these systems were illustrated with great details. We also study the poles motion of the Green's function for arbitrary (complex) frequency. Our analytic results provide a more realistic and elegant approach to study strongly correlated fermionic systems using gauge/gravity duality.

\vfill {\footnotesize Emails: zhyingfan@gmail.com }

\thispagestyle{empty}

\pagebreak



\newpage

\section{Introduction}
From our experiences in research on AdS/CFT correspondence in recent years, we can strongly believe that it provides a systematic and powerful approach to study condensed matter physics in anti-de Sitter space-time. In particular, the charged black holes with asymptotical anti-de Sitter space-time (AdS) are of great importance in applying gauge/gravity duality to study strongly interacted condensed matter systems, such as high temperature superconductors and Fermi and non-Fermi liquids. As is well-known, the Dirac-Fermi systems are first investigated in the probe limit (neglecting the back-reaction of Dirac fields) in the charged Reissner-Nordstr\"{o}m AdS black holes\cite{1}. Based on the near horizon geometry ($AdS_2\times R^{d}$), the fermionic Green's function can be half-analytically studied by performing a perturbative expansion in small $\omega$. It reads
\be G(\omega, k)=-\frac{h_1}{(k-k_F)-v_F^{-1}\omega-h_2 e^{i\gamma_{k_F}}\omega^{2\nu_{k_F}}}\label{eq1} \ee
where $k_F$ is the Fermi momentum, $v_F$ is the Fermi velocity and $h_1, h_2, \gamma_{k_F}, \nu_{k_F}$ are constants.
 A variety of Fermi liquids ($\nu_{k_F}>\frac 12$), non-Fermi liquids ($\nu_{k_F}<\frac 12$), and marginal Fermi liquids ($\nu_{k_F}=\frac 12$) are shown to be described in this model\cite{1,2,3}.

This provides an excellent example to study strongly interacted fermionic systems at finite density using gauge/gravity duality. However, the Dirac equation cannot be solved analytically and the numerical method was inevitably adopted in literatures. Although the fermionic systems having gravity duality can be adequately investigated numerically, it is more intriguing and satisfying to find analytical solutions. In fact, the Green's function $G(\omega, k)$ for massless fermions at $\omega=0$ can be found analytically in four dimensional RN-AdS black hole\cite{4}. Another progress in this direction was reported in \cite{5} in which a two-charged dilatonic black hole in $AdS_5$ was considered. It was shown in \cite{5} that in the extremal limit the Dirac equation can be solved exactly for massless fermions at $\omega=0$. The Green's function $G(0, k)$ was analytically expressed in terms of hypergeometric functions. Moreover, the Fermi momentum defined by the poles of the normal modes was also determined analytically. The perturbation method given in \cite{2,3} can then be used to obtain the analytical expressions of the constants in the Green's function in eq.(\ref{eq1}). Unfortunately, for generic frequency, there is no analytical result for the Green's function $G(\omega, k)$. Furthermore, there also lacks analytical solutions in the non-extremal black holes. Interestingly, these two problems are resolved for charged AdS black holes in conformal gravity\cite{6,7}. The Dirac equation can be solved in terms of Heun's functions for generic frequency $\omega$ and momentum $k$ which allows us to extract the Green's function $G(\omega, k)$ analytically for both extremal and non-extremal black holes.

 However, the dual boundary system has non-vanishing entropy in the extremal limit which implies that the ground state is highly degenerated. This is unfavorite in practice. As a matter of fact, the IR scaling exponent $\nu_k\propto \sqrt{k^2-k_o^2}$ ($k_o\propto q$ for massless fermion) will become imaginary for sufficiently large charge $q$\cite{8}. Once $\nu_k$ becomes imaginary, the system will be unstable by the pair production near the black hole horizon and the IR geometry will be altered to a Lifshitz geometry from the back-reaction of the pair production\cite{3,9,10,11,12}. These were particularly analyzed in Einstein gravity and we strongly believe that the qualitative features will not change in conformal gravity.

 From application of AdS/CFT correspondence to study condensed matter physics, it is more important and satisfactory to find analytical solutions in the AdS black holes which share vanishing entropy in the zero temperature limit. Motivated by this idea, we construct a series of charged dilatonic black holes using Einstein-Maxwell-Dilaton (EMD) theories in this paper. These black holes approaches to deformed AdS space-time in the extremal limit. The entropy vanishes if the temperature is sent to zero. We further show that the Dirac equation and Green's function $G(\omega, k)$ can be solved exactly in terms of hypergeometric or Heun's functions in these background. The Fermi surfaces are observed in the Green's function $G(0, k)$ as sharp peaks in the plane of momentum $k$ and fermionic charge $q$. More distinguishing features are illustrated in section 4 of this paper.

The paper is organised as follows: In section 2, we apply EMD model to construct charged dilatonic black holes. In section 3, we revisit the Dirac equation and derive the equations of motion. In section 4, we analytically solve the equations and the Green's functions. The properties of the Green's function are studied with much detail. In section 5, we shortly summarize our results.

\section{Gravity model}
In order or search charged AdS black holes with vanishing entropy in the extremal limit, we first notice that the metric and the thermal factor of the horizon can be chosen as the following form
\be ds^2=\Omega(u)^2 (-u^2 h(u)dt^2+\frac{du^2}{u^2h(u)}+u^2(dx^2+dy^2)) ,\quad h(u)=1-(\frac{u_h}{u})^{\delta}\label{eq2}\ee

where $\delta$ is a positive constant and $u_h$ defines the location of the horizon by $h(u)\equiv 0$. $\Omega(u)^2$ is an additional conformal factor which is irrelevant to determine the physics of massless spinors in the bulk since the equation of motion is conformal in this case (see eq.(\ref{eq20}-\ref{eq24}) in the next section). This allows us to freely choose the precise form of $\Omega(u)^2$. However, the full space-time should be stable which naturally requires that the null energy conditions (NEC) should be satisfied before imposing certain matter fields in the bulk. In the region far away from the horizon $u\gg u_h$, we find that

\be 2 \frac{W'^2}{W^2}-\frac{W''}{W}-2\frac{W'}{u W}\geq 0 \label{eq3}\ee

where $W(u)=u\Omega(u)$. This provides a necessary condition which we are mostly concerned about in constructing the charged gravitational background. Something else is that the metric ansatz eq.(\ref{eq2}) is identical to the dynamical holographic QCD model and its generalization\cite{13,14} except that we work in four dimensional space-time. In AdS/QCD model, one requires that the heavy quark potential is linear and the meson spectrum has linear Regge behavior.
This also provides a necessary condition on the conformal factor. Therefore, our working strategy can be parallelly discussed along the way of AdS/QCD model.

\begin{figure}[tbp]
\begin{center}
\includegraphics[width=8cm]{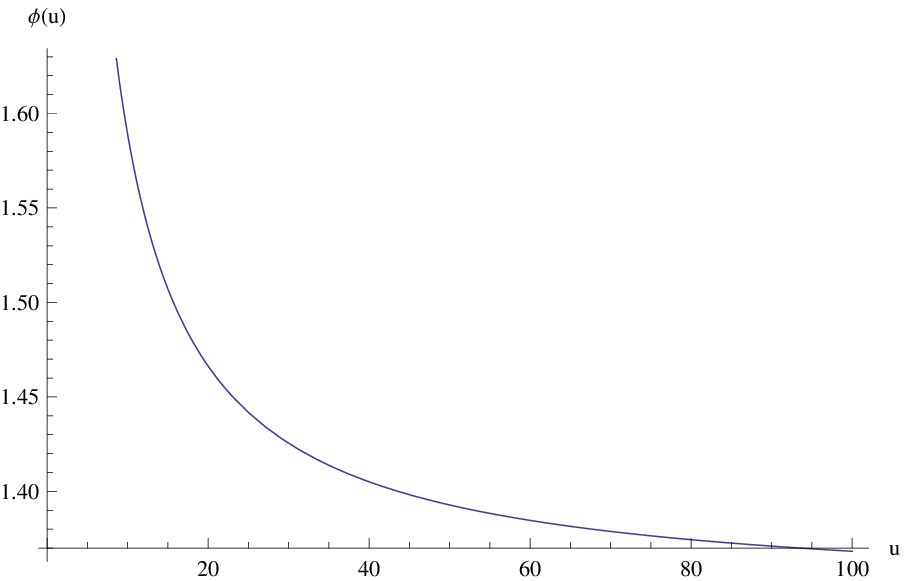}
\end{center}
\includegraphics[width=7cm]{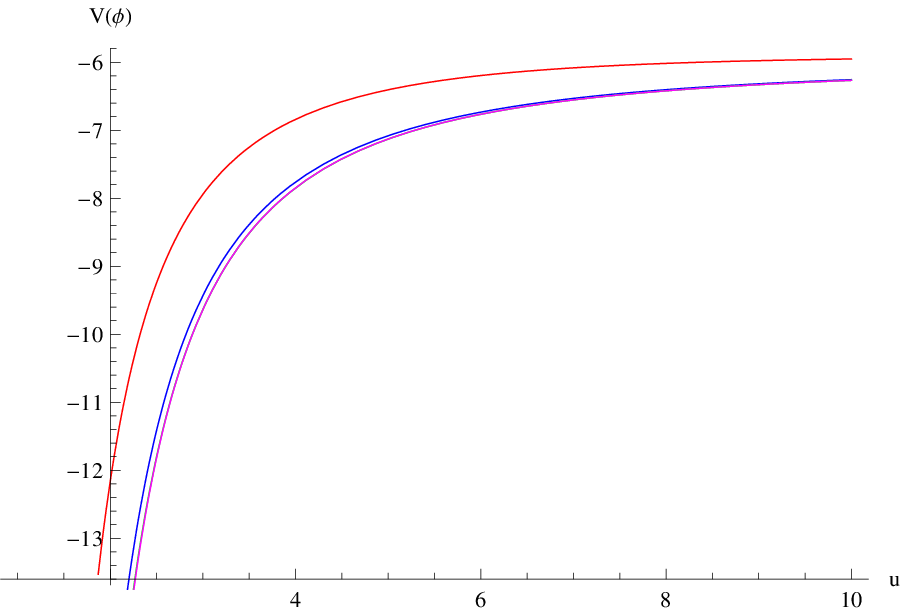}
\includegraphics[width=7cm]{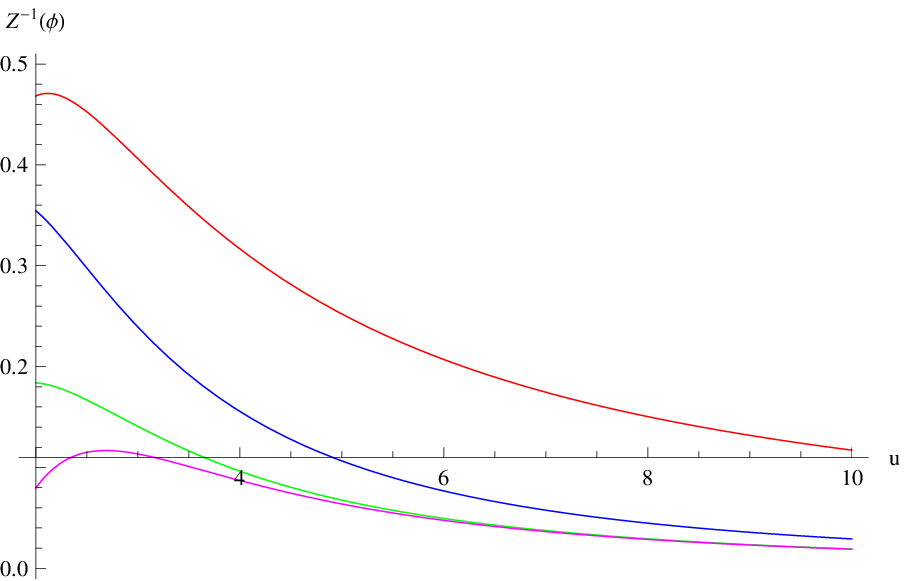}
\caption{The plots for the dilaton $\phi$, the potential $V(\phi)$ and the effective gauge coupling $Z^{-1}(\phi)$ for $\Omega(u)=e^{-1/u^2}$. In all the panels of $V(\phi)$ and $Z^{-1}(\phi)$, $\delta=1$ (red),  $\delta=2$ (blue),  $\delta=3$ (green),  $\delta=4$ (magenta). }
\label{fig1}\end{figure}

We start from the EMD action below
\be S=\frac{1}{2\kappa^2}\int \mathrm{d}^{4}x \sqrt{-g}\ [R-\frac{1}{2}F^2-2(\partial{\phi})^2-V(\phi)-\frac{1}{2}Z(\phi)H^2] \label{eq4}\ee
where $F=dA,\ H=dB$ are two U(1) gauge fields. We search static solutions for matter fields
\be \phi=\phi(u),\ \ A=A_t(u)dt,\ \ B=B_t(u)dt. \label{eq5}\ee
From Maxwell equations
\bea \partial_\mu(\sqrt{-g}F^{\mu\nu})=0, \nn\\
      \partial_\mu(\sqrt{-g}Z(\phi)H^{\mu\nu})=0,
\label{eq6}\eea
we first obtain
\be A_t(u)=\mu (1-\frac{u_h}{u}),\quad B_t'(u)=\frac{Q}{u^2Z(\phi)},\label{eq7}\ee
where $\mu$ is the chemical potential associated with the gauge field $A$ and $Q$ is the charge carried by the black holes associated with the gauge field $B$. The uu, tt, xx components of Einstein equations are listed as follows:
\be -\phi'^2+\frac{W(r)^2}{2u^4h}V(\phi)+\frac{A_t'^2}{2hW^2}+\frac{1}{2hW^2}Z(\phi)B_t'^2+3\frac{W'^2}{W^2}+\frac{W'h'}{Wh}=0,\nn\ee
\be u^4 h^2(\frac{W'^2}{W^2}-2\frac{W''}{W}-\phi'^2)
         -\frac{hW^2}{2}V(\phi)-\frac{u^4h}{2W^2}A_t'^2-\frac{u^4h}{2W^2}Z(\phi)B_t'^2-(u^4hh'+4u^3h^2)\frac{W'}{W}=0,
\nn\ee
\bea \label{eq8}u^4h(\frac{h''}{2h}+\frac{W''}{W}-\frac{W'^2}{W^2}+\phi'^2)+\frac{W^2}{2}V(\phi)\nn\\
-\frac{u^4}{2W^2}A_t'^2-\frac{u^4}{2W^2}Z(\phi)B_t'^2&&+2(u^4h'+2u^3h)\frac{W'}{W}+u^3h'=0.\eea
In addition, the equation of motion of the dilaton field is given by
\be \phi''+(\frac{h'}{h}+2\frac{W'}{W}+\frac 2u)\phi'-\frac{W^2}{4u^4h}(\frac{\partial V}{\partial \phi}+\frac 12\frac{\partial Z}{\partial \phi}H^2)=0.\label{eq9}\ee
From above equations of motion, we can obtain
\be \phi'(u)^2=2 \frac{W'^2}{W^2}-\frac{W''}{W}-2\frac{W'}{uW}, \label{eq10}\ee
\be V(\phi)=-\frac{u^4h}{W^2}(\frac{h''}{2h}+3\frac{h' W'}{h W}+\frac{W''}{W}+2\frac{W'^2}{W^2}+\frac{h'}{uh}+4\frac{W'}{uW}),\label{eq11}\ee
\be Z(\phi)=\frac{Q^2}{-A_t'^2+W(\frac 12 W h''-h W''+W'h'+\frac{h'}{u}W)}. \label{eq12}\ee
 We first note that the stability of the system requires $\phi'(u)^2\geq 0$ which is equivalent to the NEC conditions given in eq.(\ref{eq3}). Once the conformal factor $\Omega(u)$ is given, we can find the matter fields solutions from eq.(\ref{eq10}-\ref{eq12}) for the dilatonic background eq.(\ref{eq2}). In fig.\ref{fig1}, we show the dilaton potential $V(\phi)$ and the effective gauge coupling $Z^{-1}(\phi)$ for the conformal factor: $\Omega(u)=e^{-1/u^2}$. The plots are given for certain values of $\delta$ in which the Dirac equation will be solvable and thus these are the most interesting cases which will be discussed in section 4. We observe that both the dilaton potential and the effective gauge coupling converge to some constants in the UV limit $u\rightarrow \infty$. In particular, $V(\phi)\rightarrow -6$ is a universal result for generic $\delta$ in four dimensional AdS space-time. This is naturally expected since $V(\phi)$ should approaches to $2 \Lambda$, where $\Lambda$ is the cosmological constant. We can also plot $V(\phi)$ and $Z^{-1}(\phi)$ as a function of $\phi$ but in general there are no analytical expressions.

The temperature and the entropy density of the dilatonic black holes are given by:
\be T=\frac{\delta}{4\pi}u_h,\qquad s=\Omega_h^2 u_h^2,\label{eq13}\ee
where $\Omega_h\equiv \Omega(u_h)$. In the zero temperature limit $u_h\rightarrow 0$, the entropy density $s\rightarrow 0$ if $\Omega_h\rightarrow \mbox{const}$. In fact, taking this limit, the black holes eq.(\ref{eq2}) now become deformed AdS of which the deformed factor is just given by $\Omega(u)$.

To end this section, let us do a coordinate transformation:\ $r=1/u,r_h=1/u_h$ which will be more convenient for us to work when we solve the fermionic wave equations. The fermions we will consider are charged under the gauge field A which now reads: $A_t(r)=\mu(1-r/r_h)$. The metric is summarized as follows:
 \be ds^2=\frac{\Omega(r)^2}{r^2} (- h(r)dt^2+\frac{dr^2}{h(r)}+dx^2+dy^2) ,\quad h(r)=1-(\frac{r}{r_h})^{\delta}\label{eq14}\ee

\section{Dirac equation revisited}
In this section, we will derive a set of equations of motion from the Dirac equation. We just follow the procedure outlined in \cite{2,3,5}.
The Dirac equation reads:
\be  (\Gamma^a\mathcal{D}_a-M)\Psi=0 \label{eq15}\ee
where $\Gamma^a$ are the $4$ dimensional gamma matrices, $\mathcal{D}_a=(e_a)^\mu D_\mu$, with $D_\mu=\partial_\mu-i q A_\mu+\frac 14 \omega_{\mu ab}\Gamma^{ab}$, $\Gamma^{ab}=\frac 12 [\Gamma^a, \Gamma^b]$. $\omega_{\mu ab}$ is the spin connection 1-form and $M$ is the fermion mass. $(e_a)^\mu$ are vielbeins which can be chosen by
\be (e_a)^\mu=\sqrt{|g^{\mu\mu}|}(\frac{\partial}{\partial x^\mu})^a. \label{eq16}\ee
The equation of motion will be simplified by redefining the spinor as $\Psi=(-gg^{rr})^{-\frac 14}\psi$, which will remove the spin connection from the equation of motion. Due to rotational invariance, the momentum can be chosen along x direction and the Fourier mode can be set by $\psi(r,t,x)=\psi(r,\omega,k)e^{-i\omega t+ikx}$. We choose the following Gamma matrices
\begin{equation} \Gamma^r=
\left( \begin{array} {ccc}
   \sigma^3 & 0\\
   0 & \sigma^3
\end{array}\right)
\ ,\quad
\Gamma^t=
\left( \begin{array}{ccc}
  i\sigma^1 & 0 \\
  0 & i\sigma^1
\end{array}\right)
\ ,\quad
\Gamma^{x}=
\left( \begin{array}{ccc}
    -\sigma^2 & 0 \\
    0 & \sigma^2
\end{array}\right)\ ,
\nn\end{equation}
The eq.(\ref{eq15}) reduces to two decoupled equations
\be [\sqrt{g^{rr}}\sigma_3\partial_r+\sqrt{-g^{tt}}\sigma_1(\omega+q A_t)+(-1)^{\alpha}\sqrt{g^{xx}}i\sigma_2k-M]\psi_\alpha=0,\quad \alpha=1,2,  \label{eq17}\ee
where $\psi=(\psi_1, \psi_2)^T,\ \psi_1\ \mbox{and}\ \psi_2$ are two component spinors. Near the asymptotical AdS boundary $r\rightarrow 0$, $\psi_\alpha$ behaves as
\be \psi_\alpha\rightarrow a_\alpha r^M\left(\begin{array}{cc} 1\\0\end{array}\right) + b_\alpha r^{-M}\left(\begin{array}{cc}0\\1\end{array}\right)\label{eq18}\ee
If we use the standard quantization, the dual spinorial operator is left-handed and the expectation value can be read off by $\langle O\rangle=(0,b_1,0,b_2)$.
On the horizon, we impose the in-falling boundary conditions to obtain the retarded Green's function:
\be G=\left(\begin{array}{cccc}0&&&\\&G_1&&\\&&0&\\&&&G_2\end{array}\right),\quad G_\alpha=\frac{b_\alpha}{a_\alpha}.\label{eq19}\ee
 If we use the alternative quantization, the Green's function is $G_\alpha=-\frac{a_\alpha}{b_\alpha}$ and the dual spinorial operator is right-handed. For massless fermions $M=0$, $G_2$ is related to $G_1$ by $G_2=-G_1^{-1}$.

In the following, we shall focus on $\psi_1=(u_1, u_2)^T$. Define $u_\pm=u_1\pm i u_2$. From eq.(\ref{eq17}), we obtain
\be u_+'+\overline{\lambda}_1(r)u_+=\overline{\lambda}_2(r)u_-,\label{eq20}\ee
\be u_-'+\lambda_1(r)u_-=\lambda_2(r)u_+,\label{eq21}\ee
where
\be \lambda_1(r)=i\sqrt{\frac{\mid g^{tt} \mid}{g^{rr}}}(\omega+q A_t),\quad \lambda_2(r)=\frac{M}{\sqrt{g^{rr}}}-ik\sqrt{\frac{g^{xx}}{g^{rr}}}. \label{eq22}\ee
It is immediately seen that the equations of motion for massless fermions depend only on the ration of the metric components and hence are conformal. From eq.(\ref{eq20}-\ref{eq21}), we can further derive two decoupled second-order differential equations:
\be u_+''+\overline{p}_1(r)u_+'+\overline{p}_2(r)u_+=0, \label{eq23}\ee
\be u_-''+p_1(r)u_-'+p_2(r)u_-=0, \label{eq24}\ee
where
\be p_1(r)=-\frac{\lambda_2'}{\lambda_2},\quad p_2(r)=|\lambda_1|^2-|\lambda_2|^2+p_1\lambda_1+\lambda_1'. \label{eq25}\ee
It is also immediately seen that $u_+$ takes the form which is complex conjugate of $u_-$ from eq.(\ref{eq23}-\ref{eq24}). However, since we have only two boundary conditions, we need first solve eq.(\ref{eq24}) for $u_-$ and then plug in $u_-$ to the first order equation eq.(\ref{eq21}) to obtain the integration constant and $u_+$. Once the wave functions $u_\pm$ are obtained, the Green's function can be read off from their asymptotic behavior at the AdS boundary:
\be G_1=-G^{-1}_2=\lim_{r\rightarrow 0}\frac{u_2}{u_1}=-i\lim_{r\rightarrow 0}\frac{u_+-u_-}{u_++u_-}.\label{eq26}\ee

\section{Exact Green's function}

In this section, we will solve the equations of motion for massless fermions in the background eq.(\ref{eq14}). To simplify calculation, we will work in the rescaled radial coordinate $\tilde{r}\equiv r/r_h$ in which the horizon was scaled to unity. Correspondingly, the frequency, momentum and the effective chemical potential are also rescaled as: $\tilde{\omega}=\omega r_h, \tilde{k}=k r_h,\tilde{\mu}_q=\mu_q r_h$, where $\mu_q\equiv \mu q$. From eq.(\ref{eq13}), all the dependence on the temperature in the wave functions and the Green's functions is encoded in the horizon radius $r_h$. Since no true extremal limit exists in this background, the Green's functions are actually always solved at finite temperature. This is qualitatively different from published works in conformal gravity\cite{6,7} in which the extremal limit can be taken as well as those in RN black holes. As a matter of fact, the poles in the Green's function moves in a different way in the complex frequency plane, as will be shown in the following.

\subsection{Case 1: $\delta=1$}
\begin{figure}[tbp]
\includegraphics[width=7cm]{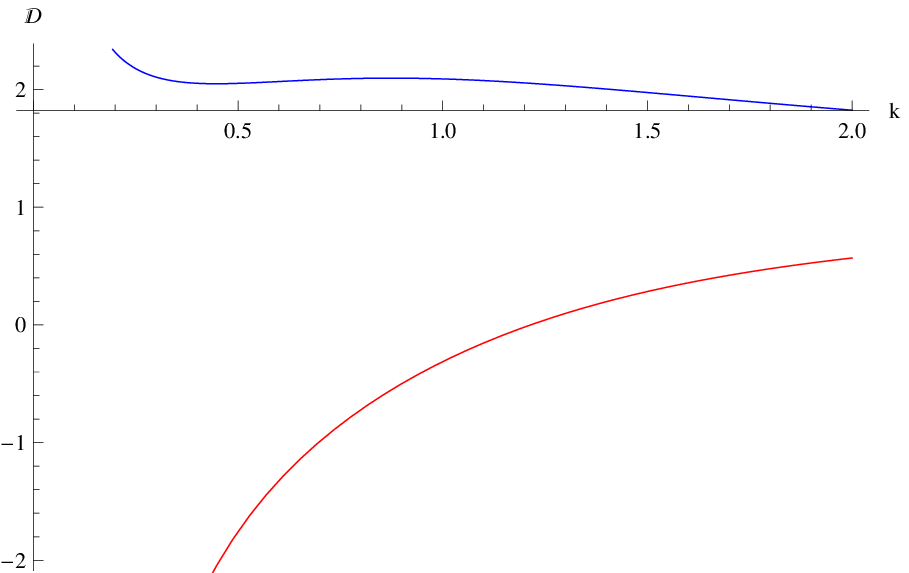}
\includegraphics[width=7cm]{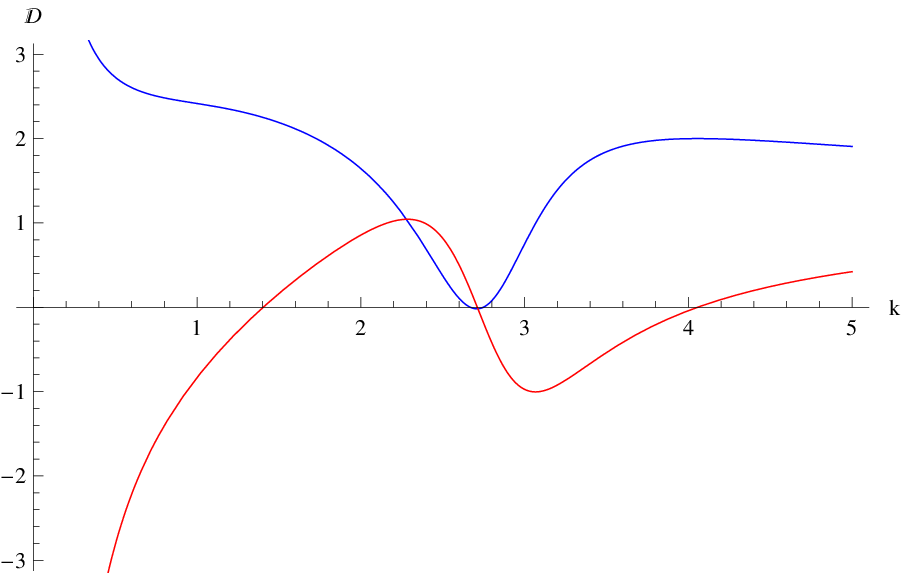}
\caption{These are the plots of the real (blue lines) and imaginary (red lines) part of $\mathcal{D}(0,k)$ as functions of k for fixed q. In the left plot, q=2 and no Fermi surfaces exist; in the right plot, q=5 and a Fermi surface emerges at $k_F=2.71134486$.  }
\label{fig2}\end{figure}
This is the simplest case in which the Dirac equation can be exactly solved. The general solutions are Whittaker functions. We choose only the in-falling solutions for $u_\pm$:
\be u_-(r)=(1-\tilde{r})^{-\frac 14}M_{\lambda,\nu}(z),\qquad u_+(r)=c (1-\tilde{r})^{-\frac 14}M_{\lambda-\frac 12,\nu-\frac 12}(z), \label{eq27}\ee
where $M_{\lambda,\nu}(z)$ is defined by the confluent hypergeometric function:
\be M_{\lambda,\nu}(z)=z^{\nu+\frac 12}e^{-z/2} {}_1F_1(\nu-\lambda+\frac 12, 2\nu+1;z) \label{eq28}\ee
Further notations are specified by:
\be\lambda=\frac 14+i\tilde{\omega}-i\frac{\tilde{k}^2}{2\tilde{\mu}_q},\quad \nu=\frac 14-i\tilde{\omega},\quad c=2(1+i)\frac{\nu\sqrt{\tilde{\mu}_q}}{\tilde{k}},\quad z=2i\tilde{\mu}_q(\tilde{r}-1).\label{eq29}\ee
As $\tilde{r}$ approaches to the horizon, we have $z\rightarrow 0$ and the Whittaker function behaves as $M_{\lambda,\nu}(z)\sim z^{\nu+\frac 12}$. Thus, the solutions given by eq.(\ref{eq27}) are indeed in-going near the horizon. We can now immediately write down the retarded Green's function from eq.(\ref{eq26}):
\be G(\omega,k)=i\frac{1-\mathcal{R}(\omega,k)}{1+\mathcal{R}(\omega,k)},\qquad
\mathcal{R}(\omega,k)=c\frac{M_{\lambda-\frac 12,\nu-\frac 12}(-2i\tilde{\mu}_q)}{M_{\lambda,\nu}(-2i\tilde{\mu}_q)}.\label{eq30}\ee
Here by $G(\omega,k)$ we mean $G_1(\omega,k)$. The $G_2$ component can be obtained by its inverse. The Fermi surfaces are defined by the poles or zeros (for alternative quantization) of the Green's function at $\omega=0$. For certain choices of $(k,q)$, the Green's function $G(0,k)$ diverges or vanishes. Then Fermi surfaces emerge at such points in the $(k,q)$ plane. In the following, we will focus on discussing the standard quantization. It is straightforward to study the alternative quantization in a parallel way.

Since $G(0,k)$ is in general complex, the definition of Fermi surfaces actually requires that both the real and imaginary part of its denominator vanishes. This will generally give a ``complex Fermi momentum" which has a very small imaginary part. In practice, we shall drop the imaginary part. The Green's function then becomes sufficiently large in many orders of magnitude instead of diverging at the Fermi momentum literally. This shall be regarded as a practical definition for Fermi momentum and Fermi surfaces. Unfortunately, we cannot deduce an analytical expression for the Fermi momentum although we have solved the Green's function exactly. Nevertheless, we will present illustrative examples to examine the denominator of the Green's function eq.(\ref{eq30}):
\be \mathcal{D}(\omega,k)=1+\mathcal{R}(\omega,k).\label{eq31}\ee
The zeros of $\mathcal{D}(0,k)$ lead to the Fermi momenta. Since $\mathcal{D}$ is in general a complex number, it is unlikely that both the real and imaginary parts of $\mathcal{D}$ vanish strictly for certain values of $(k,q)$. Thus, the Fermi momentums we are searching for are those values of momentum at which the minimum of the denominator is sufficiently small in many orders of magnitude.
\begin{figure}[tbp]
\includegraphics[width=7cm]{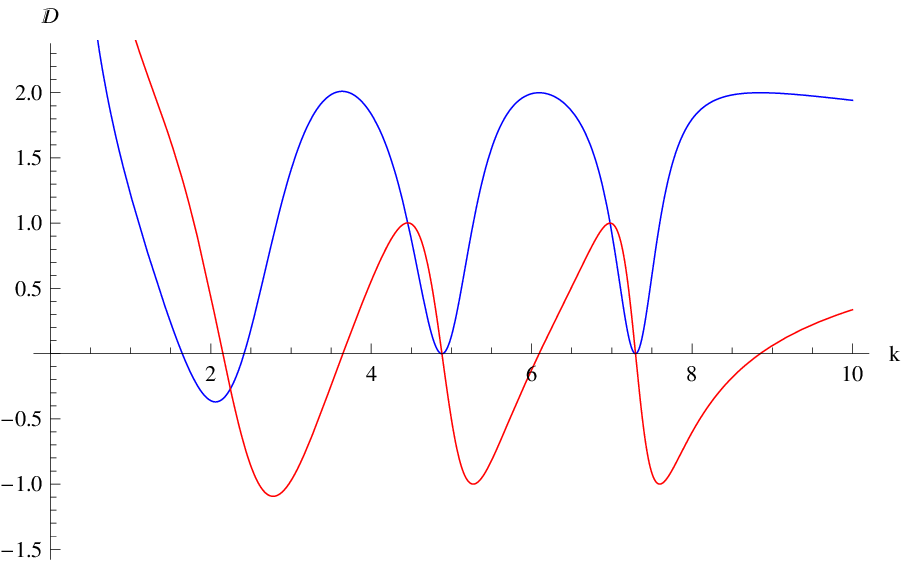}
\includegraphics[width=7cm]{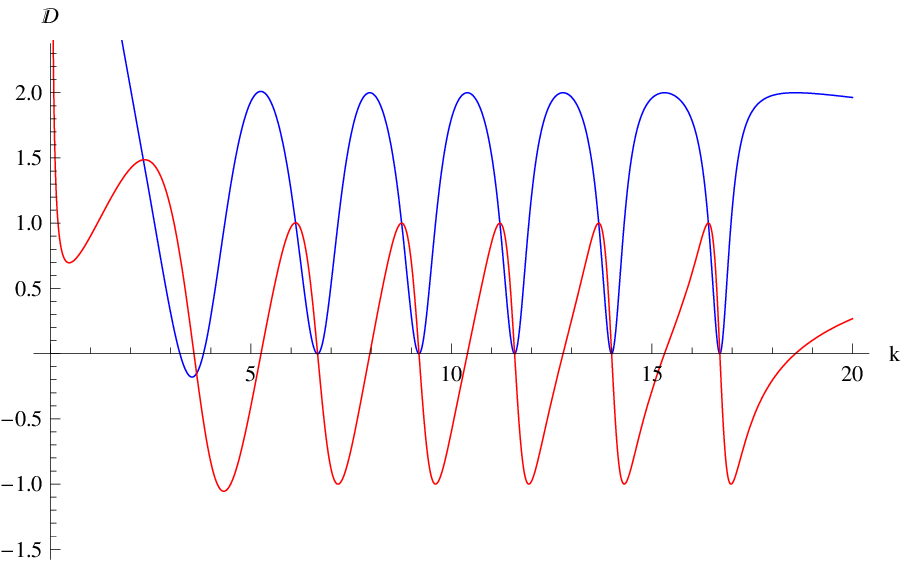}
\caption{These are the plots of the real (blue lines) and imaginary (red lines) part of $\mathcal{D}(0,k)$ as functions of k for fixed q. In the left plot, q=10 and two Fermi surfaces occur at $k_1=4.88140054,\ k_2=7.29508109$; in the right plot, q=20 and five Fermi surfaces occur at $k_1=6.66251373,\ k_2=9.19082672,\ k_3=11.57479844,\ k_4=13.99473695,\ k_5=16.69249918$.  }
\label{fig3}\end{figure}

In fig.\ref{fig2}\footnote{For all the figures given in this section, we have set the horizon $r_h=1$ and the chemical potential $\mu=1$.}, we show the real and imaginary part of $\mathcal{D}(0,k)$ as a function of k for some fixed q. The left plot is given for $q=2$. It is evident that the real and imaginary part do not approach zero simultaneously as k varies. Therefore, no Fermi surfaces exist in this case. This is a generic feature for small q which can be verified from eq.(\ref{eq30}) by sending $q\rightarrow 0$. However, when q becomes larger, the situation changes. In the right plot of fig.\ref{fig2}, we have $q=5$ and a Fermi surface appears at $k_F=2.71134486$. As $q$ increases, more and more Fermi surfaces emerge, as is shown in fig.\ref{fig3} in which we plot the real and imaginary part of $\mathcal{D}(0,k)$ for $q=10$ and $20$.

\begin{figure}[tbp]
\includegraphics[width=7cm]{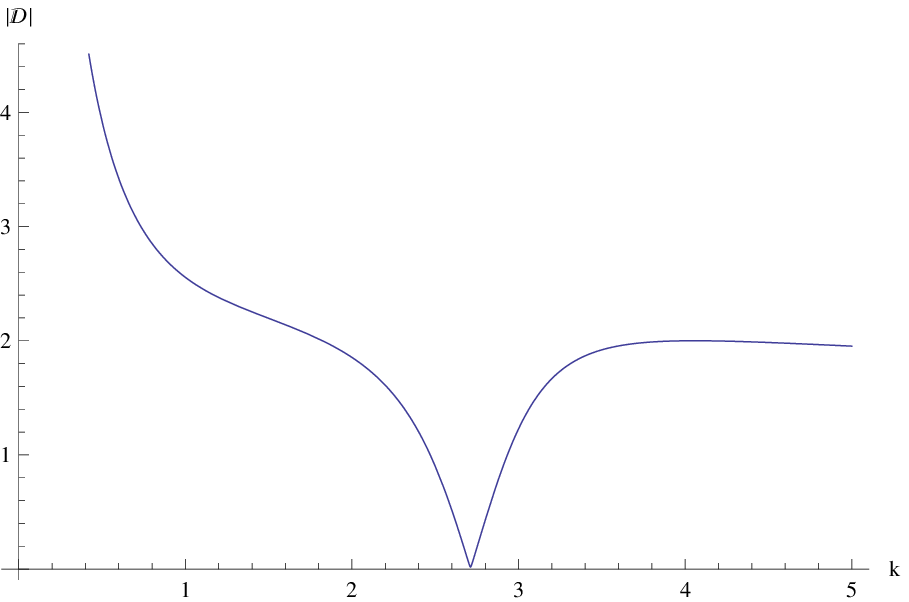}
\includegraphics[width=7cm]{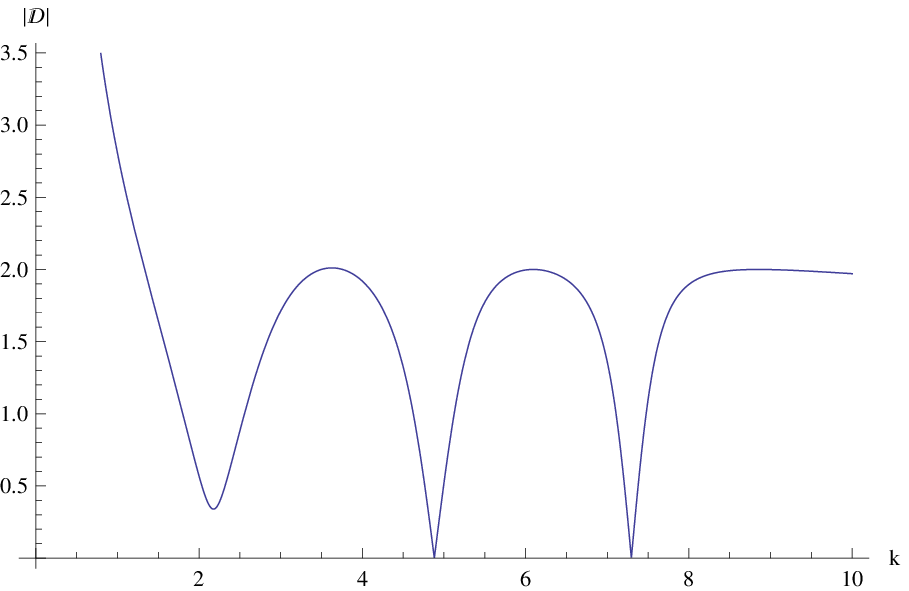}
\caption{The plots of $|\mathcal{D}(0,k)|$ for fixed q. The left plot for $q=5$; the right plot for $q=10$.  }
\label{fig4}\end{figure}

It is worth noting that in above figures, when the imaginary part (red lines) cross zeros, the real part (blue lines) some times does not approach zeros literally. Instead, it approaches a local minimum which is very small, but not necessarily zero. For example, for $q=5$, at the Fermi momentum $k_F=2.71134486$, we have $Re(\mathcal{D})=0.0181$. For $q=10$, at $k_1=4.88140054$, we have  $Re(\mathcal{D})=9.5526\times 10^{-4}$; at $k_2=7.29508109$, we have $Re(\mathcal{D})=1.3731\times 10^{-7}$. Due to the fact that $\mathcal{D}$ is complex, the absolute value $|\mathcal{D}|$ does not hit zero literally. However, its local minimums are so small that they can signal Fermi surfaces in practical purpose. In fig.\ref{fig4}, we show $|\mathcal{D}(0,k)|$ for $q=5$ and 10.

\begin{figure}[tbp]
\includegraphics[width=7cm]{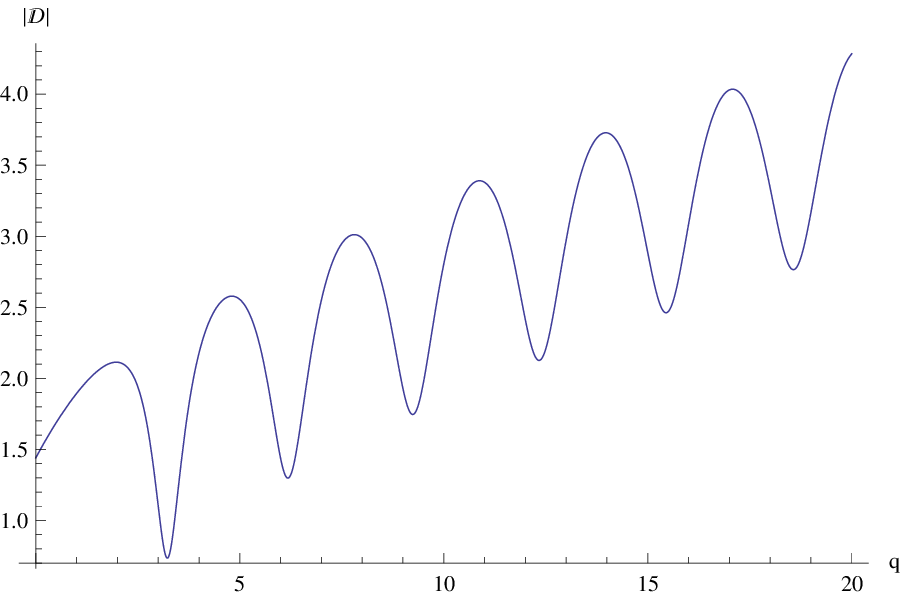}
\includegraphics[width=7cm]{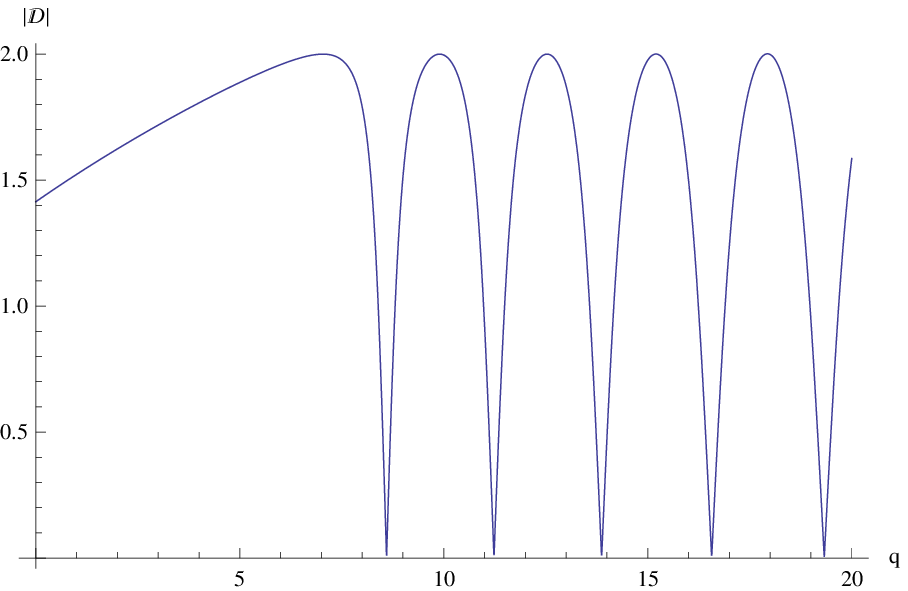}
\caption{The plots of $|\mathcal{D}(0,k)|$ for fixed k. The left plot for $k=1$; the right plot for $k=6$.  }
\label{fig5}\end{figure}

We have studied how the Fermi surfaces emerge for fixed q. It is also instructive to study the behavior of the Green's function for fixed k by varying q from 0 to infinity. In fig.\ref{fig5}, we present the graphs on the absolute value $\mathcal{D}$ as a function of q for fixed k. In the left plot, we let $k=1$ and find that the first local minimum of $|\mathcal{D}|$ is above 0.7. Hence, there are no Fermi surfaces. In the right plot, we choose $k=6$ and find that multiple Fermi surfaces emerge.

Up to now, we have studied the Green's function at $\omega=0$. It is of great interests to study how the Green's function behaves when $\omega$ varies. We first let $\omega$ to be real and show the behavior of the Green's function at a given Fermi momentum. In fig.6, we present a graph for $|G(\omega,k_F)|$ for $q=20$. We immediately observe that there are multiple maxima spikes. The $\omega=0$ spike is the Fermi surface for $k=9.19082672$ and the right spike is the Fermi surface having appeared at $k=6.66251373$. The left spikes are those Fermi surfaces having Fermi momentum larger than $9.19082672$. Therefore, in the $(k,q)$ plane with vanishing $\omega$, there are five Fermi surfaces for $q=20$ which is consistent with previous results. We conclude that when $k$ increases, every time a spike crosses the vertical axis, a Fermi surface emerges.
\begin{figure}[tbp]
\begin{center}
\includegraphics[width=8.5cm]{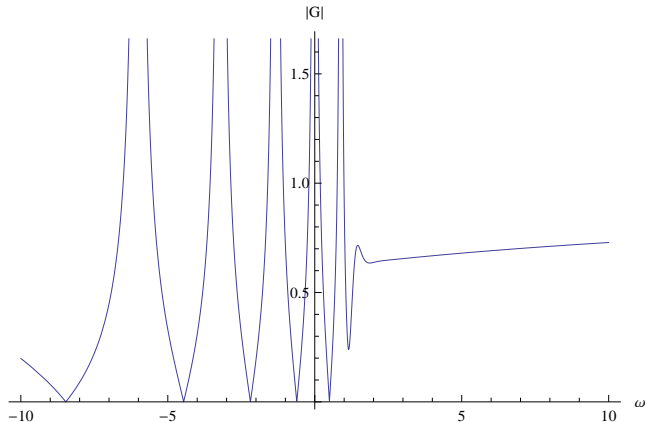}
\caption{The plots of $|G(\omega,k_F)|$ for $q=20$. The Fermi momentum is $k_F=9.19082672$. We see that multiple peaks arise when we change $\omega$. }
\end{center}
\label{fig6}\end{figure}

\begin{figure}[tbp]
\includegraphics[width=5cm]{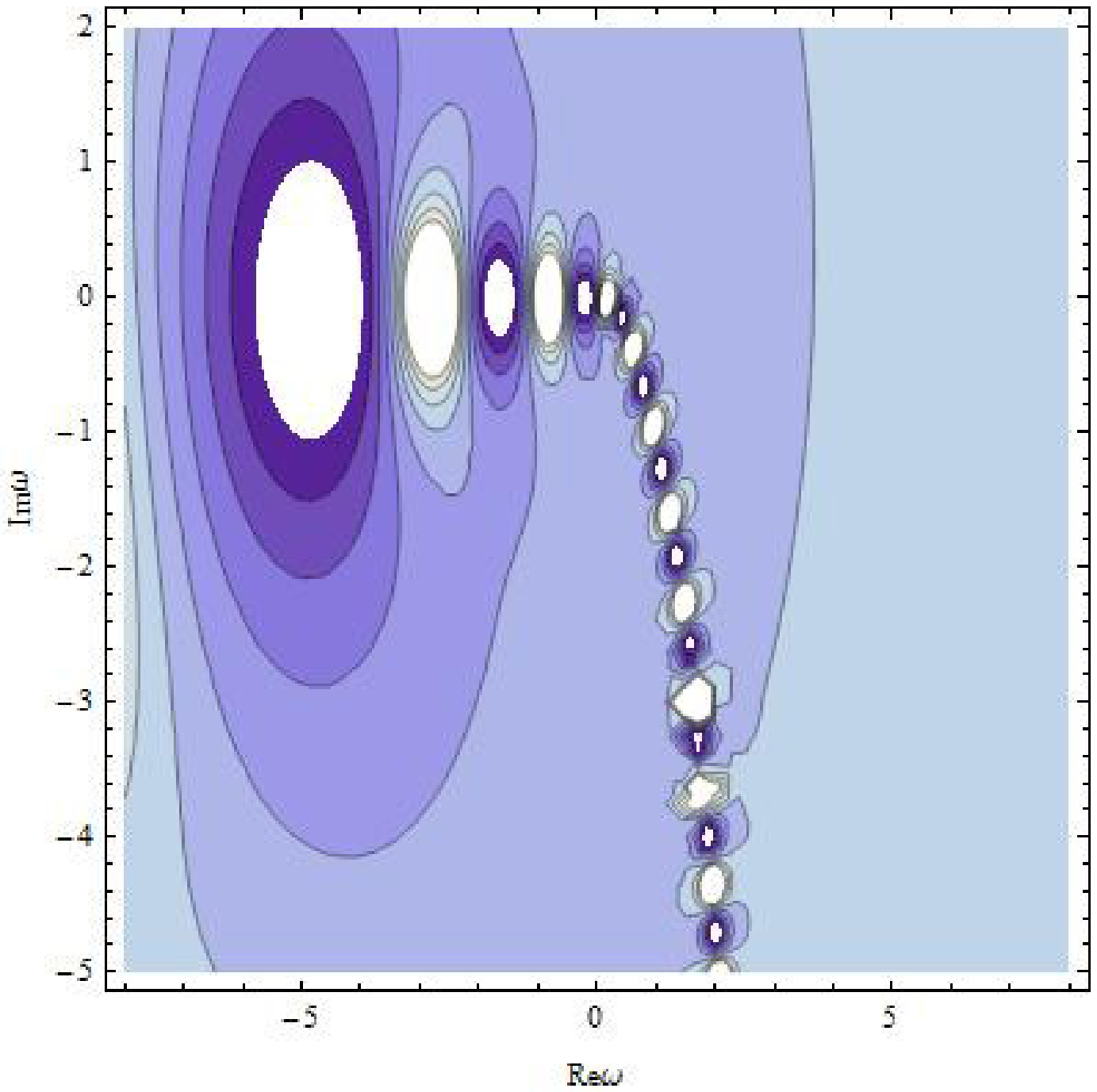}
\includegraphics[width=5cm]{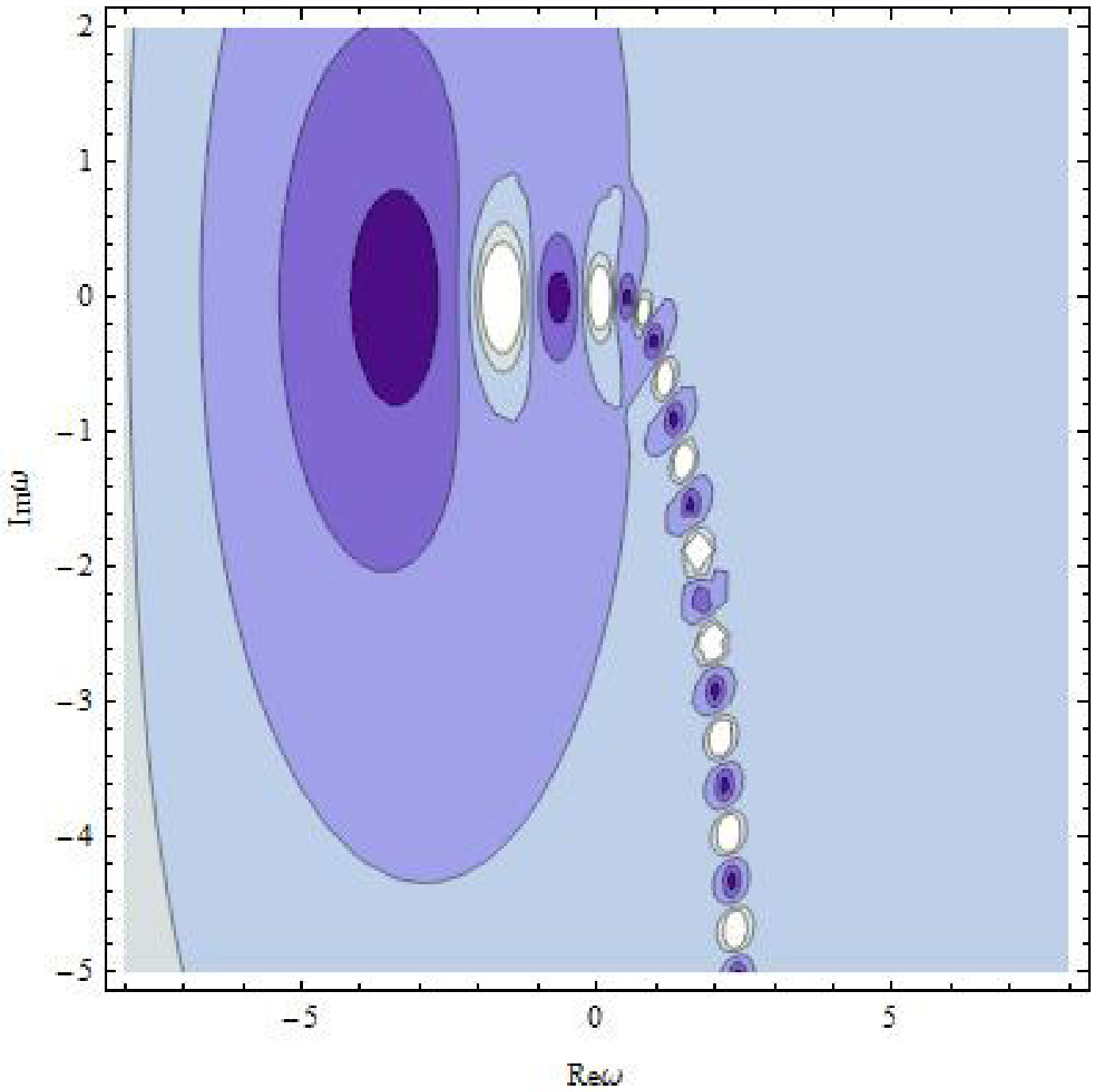}
\includegraphics[width=5cm]{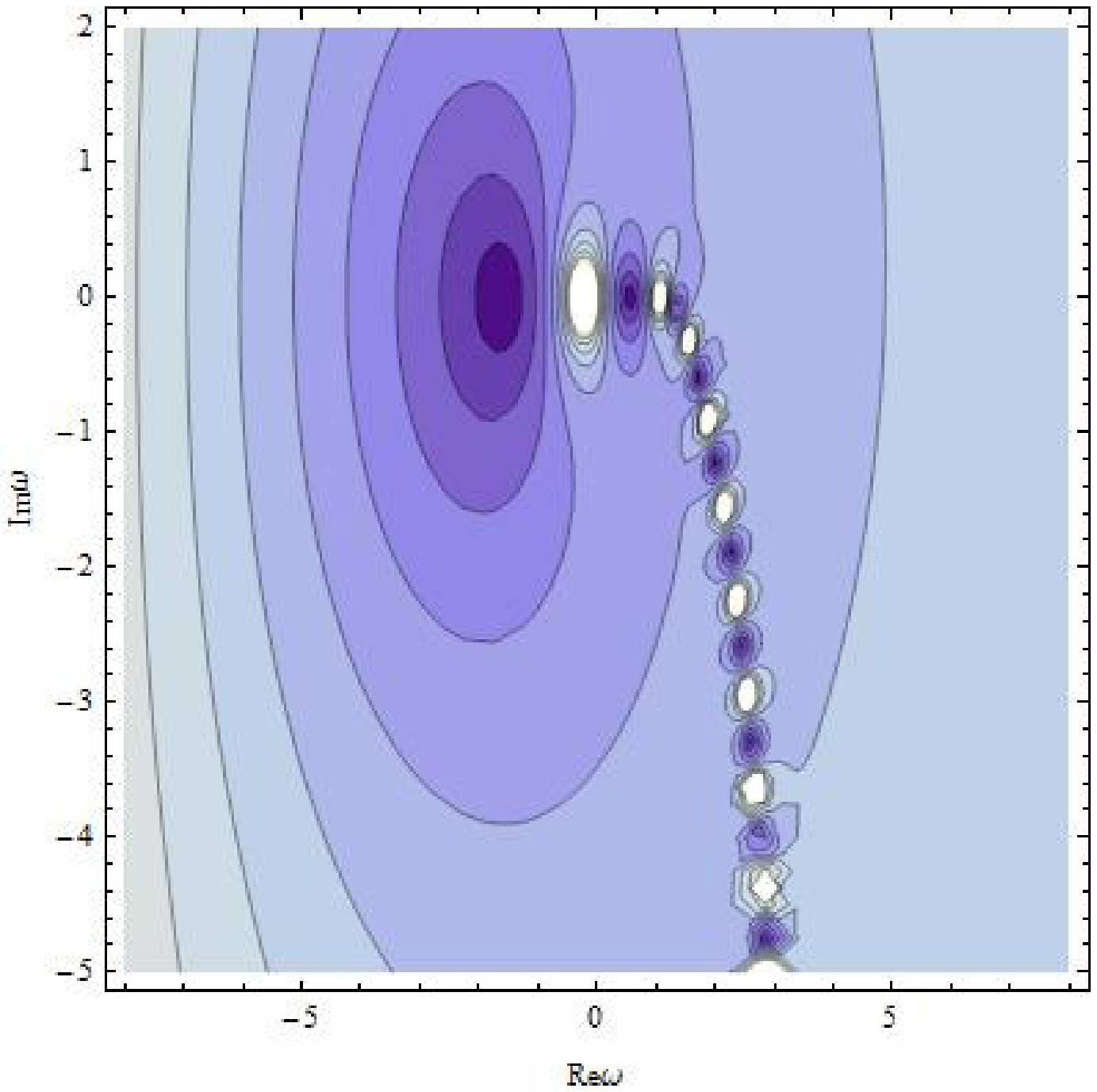}
\caption{The contour plots of $|G(\omega,k)|$ in the plane of complex $\omega$ for fixed k. The spinor charge is $q=10$. The left plot for $k=3$, the middle plot for $k=5$ and the right plot for $k=7$. All the poles (white dots) and zeros (dark dots) occur in the lower half plane.}
\label{fig7}\end{figure}

We then set $\omega$ to be complex. Since our fermions are only probe in the dilatonic black holes, the Green's function with complex frequency is therefore going to signify the features of quasi-normal modes (QNMs). The poles motion of the Green's function encode the information on fermionic stability or instability. In fig.\ref{fig7}, we present three contour plots of $G(\omega,k)$ for $q=10$ in the plane of complex frequency for certain values of $k$. We observe that as $k$ increases, the poles move to the right. As $\omega=0$ corresponds to the Fermi surface, every time a pole crosses the imaginary axis, we obtain a Fermi surface. Thus, given a sufficiently large $q$, we can obtain many Fermi momenta and many Fermi surfaces. Notice that all the poles and zeros occur in the lower half plane, indicating that our fermionic theory is stable against small perturbations.

\begin{figure}[tbp]
\includegraphics[width=5cm]{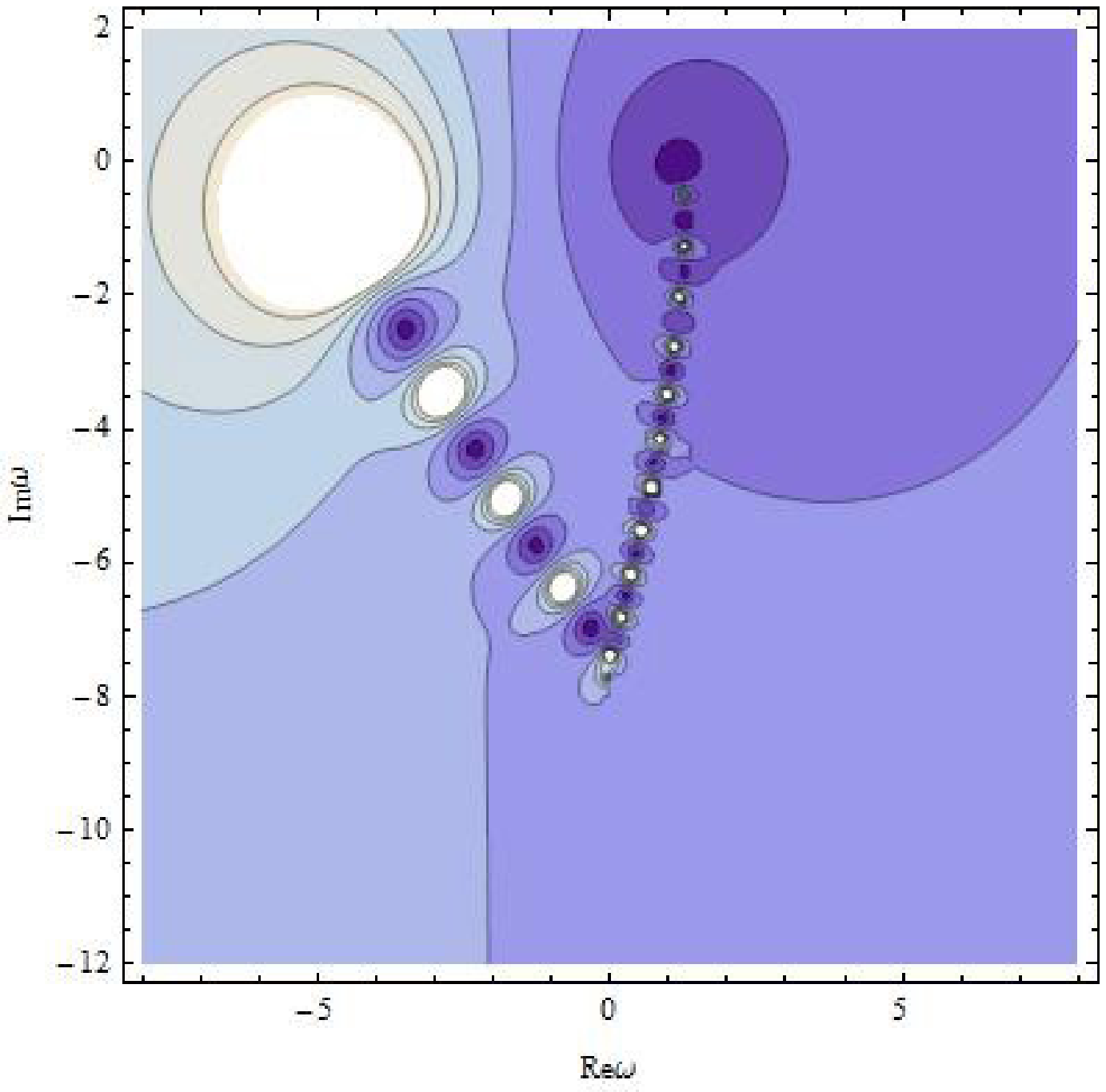}
\includegraphics[width=5cm]{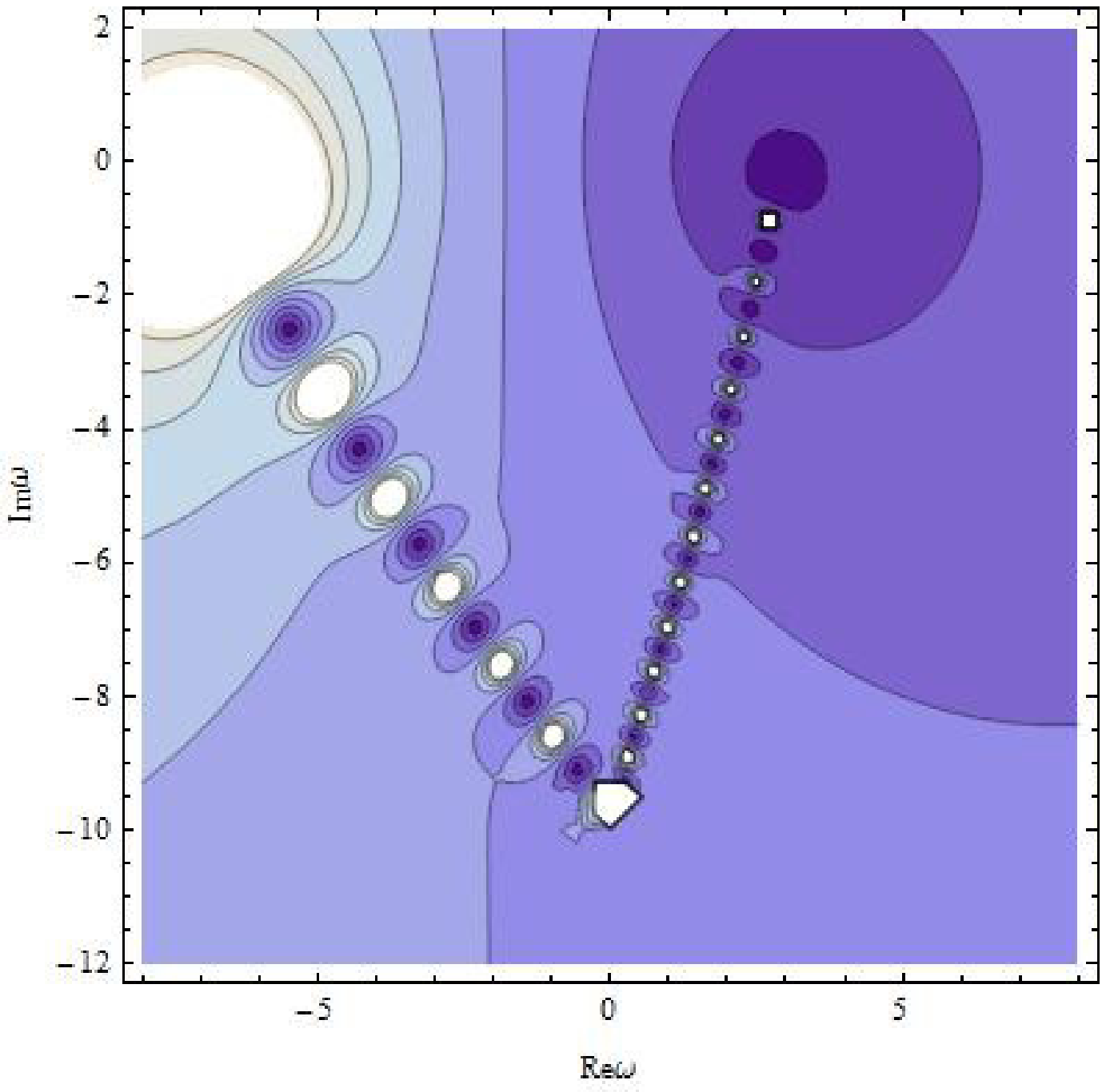}
\includegraphics[width=5cm]{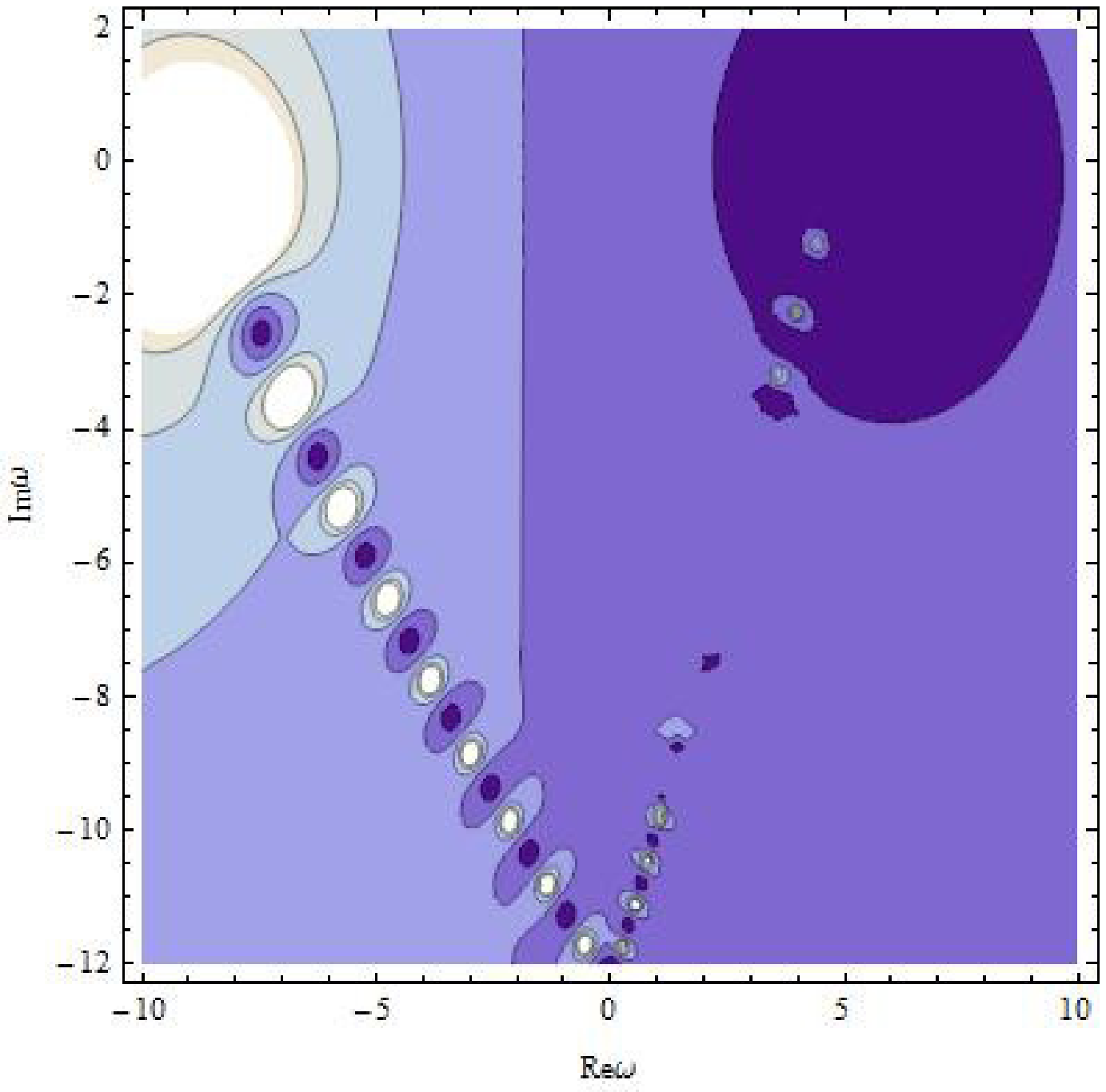}
\caption{The contour plots of $|G(\omega,k)|$ in the plane of complex $\omega$ for fixed k. The spinor charge is $q=2$. The left plot for $k=3$, the middle plot for $k=5$ and the right plot for $k=7$.  All the poles (white dots) and zeros (dark dots) occur in the lower half plane. In the right plot, the poles and zeros in the right plane becomes too small to be shown in the figure.}
\label{fig8}\end{figure}

Another example is given for $q=2$ in fig.\ref{fig8}. We find that there are two branches of poles and zeros in the plane of complex frequency, with one in the left plane and one in the right plane. The two branch joins at the imaginary axis. When we increase $k$, the number of poles and zeros in both branches increases and the joint point moves towards the lower half plane.
\subsection{Case 2: $\delta=2$}
For $\delta=2$, the wave function $u_\pm$ are solved in terms of hypergeometric functions:
\bea u_-(r)&=&(1-\tilde{r})^{\frac{1-i\tilde{\omega}}{2}}(1+\tilde{r})^{(\frac{1+i\tilde{\omega}}{2}+i\tilde{\mu}_q)}
{}_2F_1(1+i\tilde{\mu}_q+i\gamma_k,1+i\tilde{\mu}_q-i\gamma_k,\frac 32-i\tilde{\omega},\frac{1-\tilde{r}}{2}),\nn\\
 u_+(r)&=&c(1-\tilde{r})^{-\frac{i\tilde{\omega}}{2}}(1+\tilde{r})^{(\frac{i\tilde{\omega}}{2}+i\tilde{\mu}_q)}
{}_2F_1(i\tilde{\mu}_q+i\gamma_k,i\tilde{\mu}_q-i\gamma_k,\frac 12-i\tilde{\omega},\frac{1-\tilde{r}}{2}),\label{eq32}\eea
where some notations are specified by
\be \gamma_k=\sqrt{\tilde{\mu}_q^2+\tilde{k}^2},\quad c=-\frac{2\tilde{\omega}+i}{\tilde{k}}.\label{eq33}\ee

\begin{figure}[tbp]
\includegraphics[width=7cm]{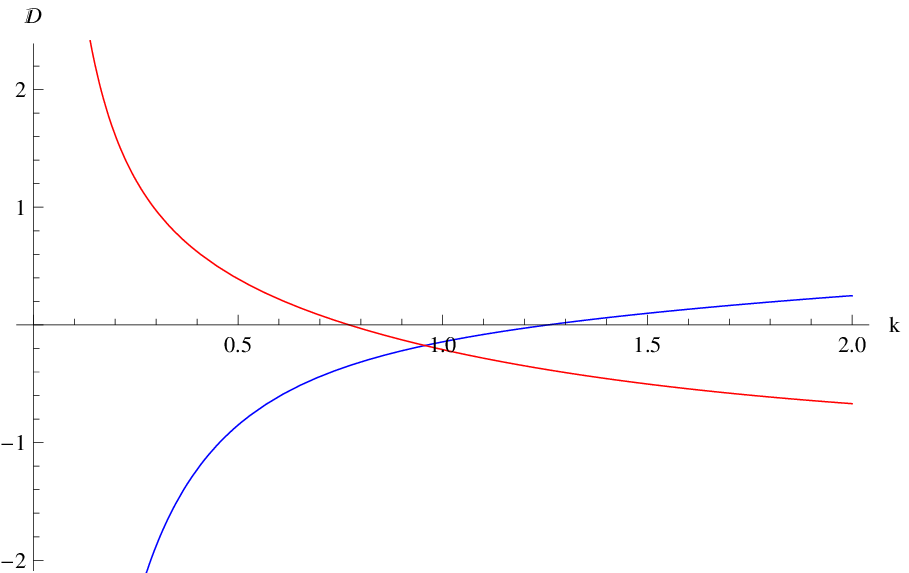}
\includegraphics[width=7cm]{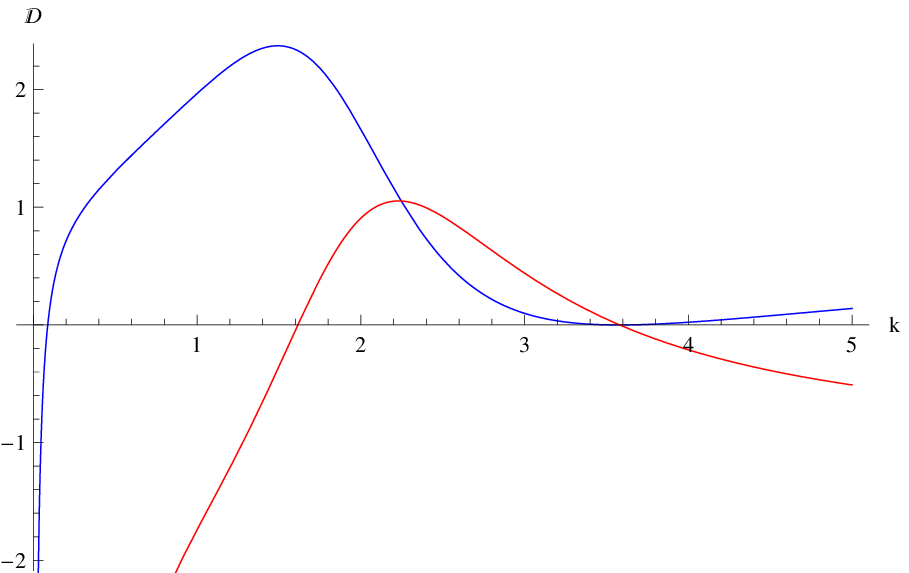}
\caption{These are the plots of the real (blue lines) and imaginary (red lines) part of $\mathcal{D}(0,k)$ as functions of k for fixed q. In the left plot, q=2 and no Fermi surfaces exist; in the right plot, q=5 and a Fermi surface emerges at $k_F=3.58013600$.  }
\label{fig9}\end{figure}
\begin{figure}[tbp]
\includegraphics[width=7cm]{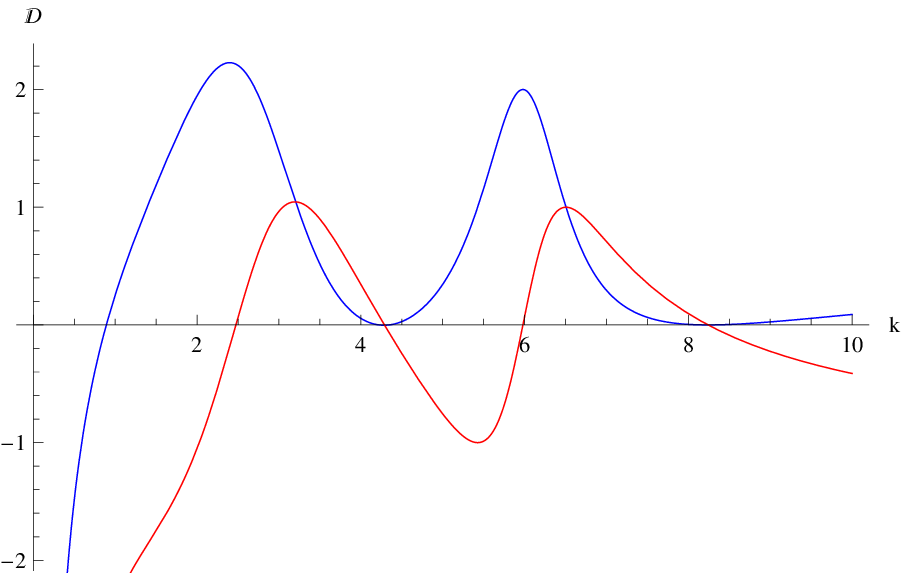}
\includegraphics[width=7cm]{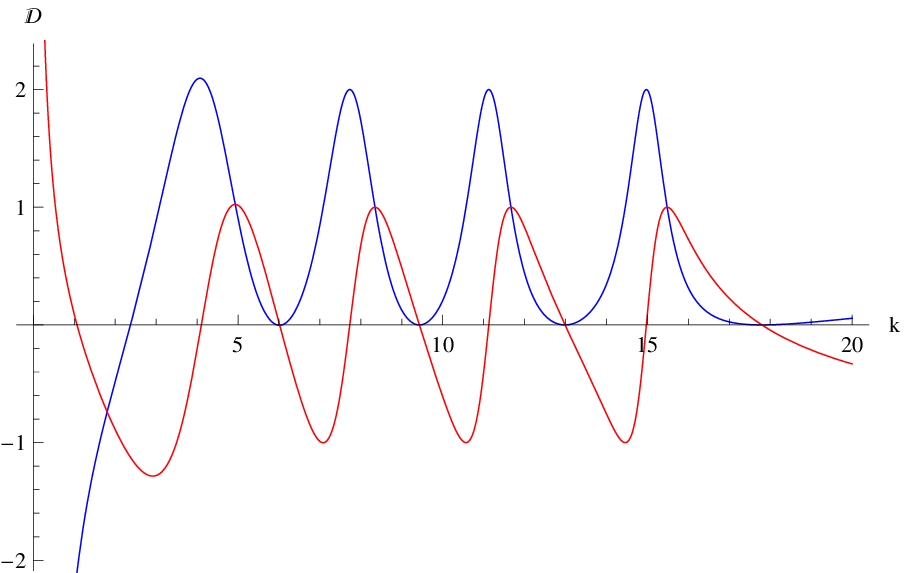}
\caption{These are the plots of the real (blue lines) and imaginary (red lines) part of $\mathcal{D}(0,k)$ as functions of k for fixed q. In the left plot, q=10 and two Fermi surfaces occur at $k_1=4.28816811,\ k_2=8.24047002$; in the right plot, q=20 and four Fermi surfaces occur at $k_1=6.01641240,\ k_2=9.42936683,\ k_3=12.98725992,\ k_4=17.79572499$.  }
\label{fig10}\end{figure}

The Green's function can be expressed as:
\be G(\omega,k)=i\frac{1-\mathcal{R}(\omega,k)}{1+\mathcal{R}(\omega,k)},\qquad
\mathcal{R}(\omega,k)=c\frac{{}_2F_1(i\tilde{\mu}_q+i\gamma_k,i\tilde{\mu}_q-i\gamma_k,\frac 12-i\tilde{\omega},\frac 12)}
{{}_2F_1(1+i\tilde{\mu}_q+i\gamma_k,1+i\tilde{\mu}_q-i\gamma_k,\frac 32-i\tilde{\omega},\frac 12)}.\label{eq34}\ee
Since the Fermi surfaces are encoded in the poles of the Green's function, we are particularly interested in studying its denominator $\mathcal{D}\equiv 1+\mathcal{R}$. First, we consider the zero modes: $\omega=0$. In fig.\ref{fig9}, we plot $\mathcal{D}(0,k)$ as functions of $k$ for fixed spinor charge $q$. In the left plot, we have $q=2$ but find no Fermi surface. In the right plot, we have $q=5$ and one Fermi surface emerges at the momenta $k_F=3.58013600$. When the charge $q$ increases, we find more and more Fermi surfaces, as is shown in fig.\ref{fig10}.

\begin{figure}[tbp]
\includegraphics[width=7cm]{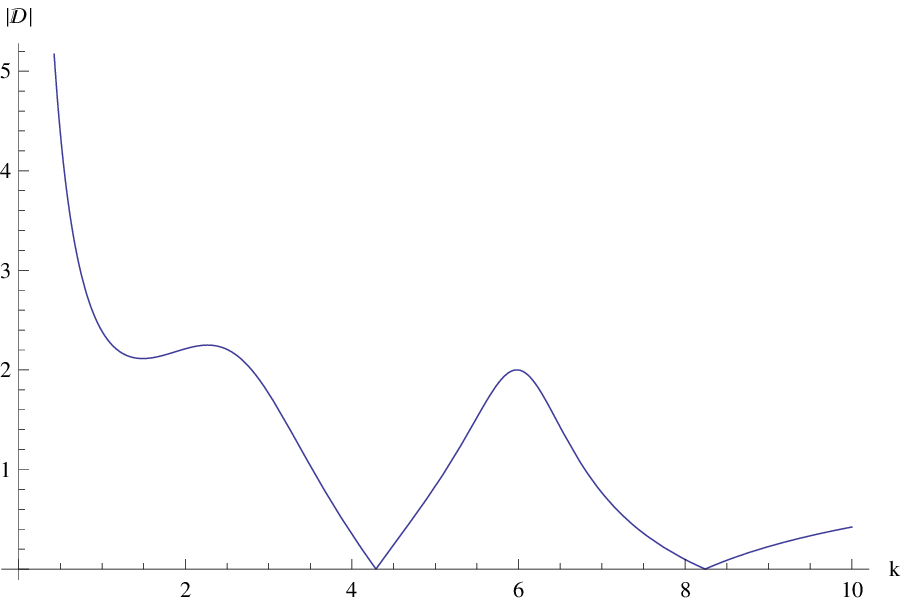}
\includegraphics[width=7cm]{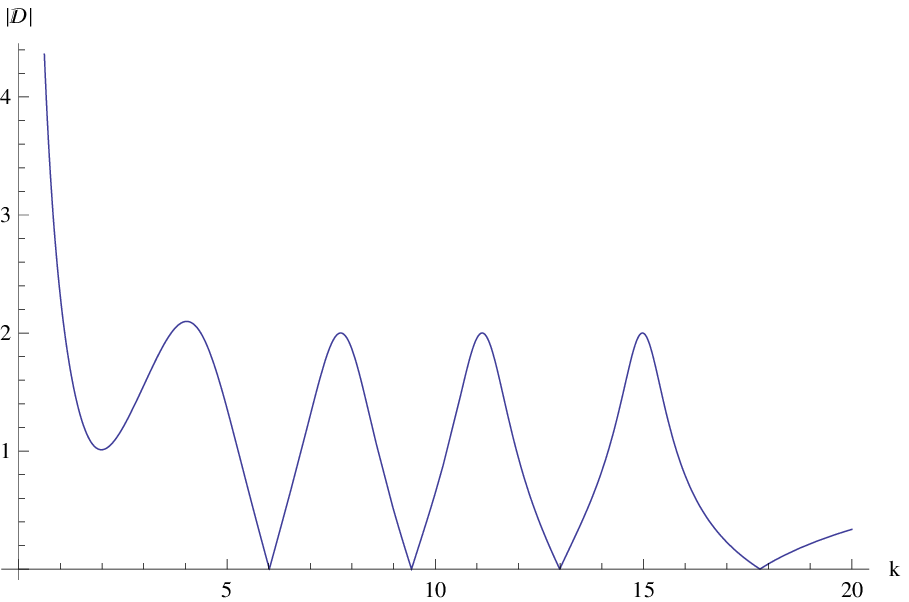}
\caption{The plots of $|\mathcal{D}(0,k)|$ for fixed q. The left plot for $q=10$; the right plot for $q=20$.  }
\label{fig11}\end{figure}

As we have discussed before, the local minimums of $|\mathcal{D}|$ are sufficient to define Fermi surfaces in practical purpose. This is also true for $\delta=2$ case. For example, when $q=5$, we have $|\mathcal{D}|=3.221\times 10^{-4}$ at the Fermi momenta $k_F=3.58013600$ and when $q=10$, we have $\mathcal{D}=0.0025$ for the first Fermi momenta $k_1=4.28816811$. In fig.\ref{fig11}, we present two examples in which $|\mathcal{D}|$ is drew for $q=10$ (the left plot) and $q=20$ (the right plot). The encoded information on Fermi surfaces is consistent with that given in fig.\ref{fig10}.
\begin{figure}[tbp]
\includegraphics[width=7cm]{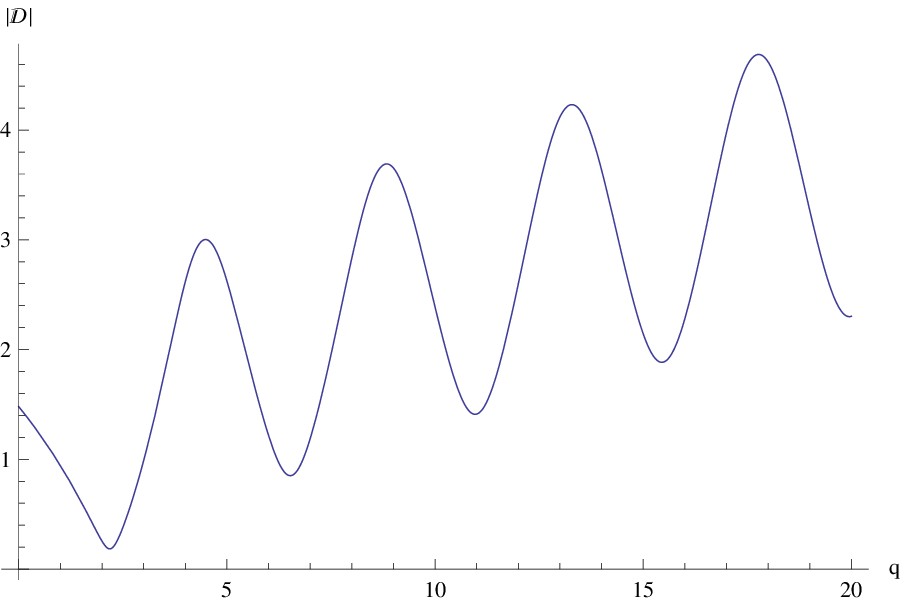}
\includegraphics[width=7cm]{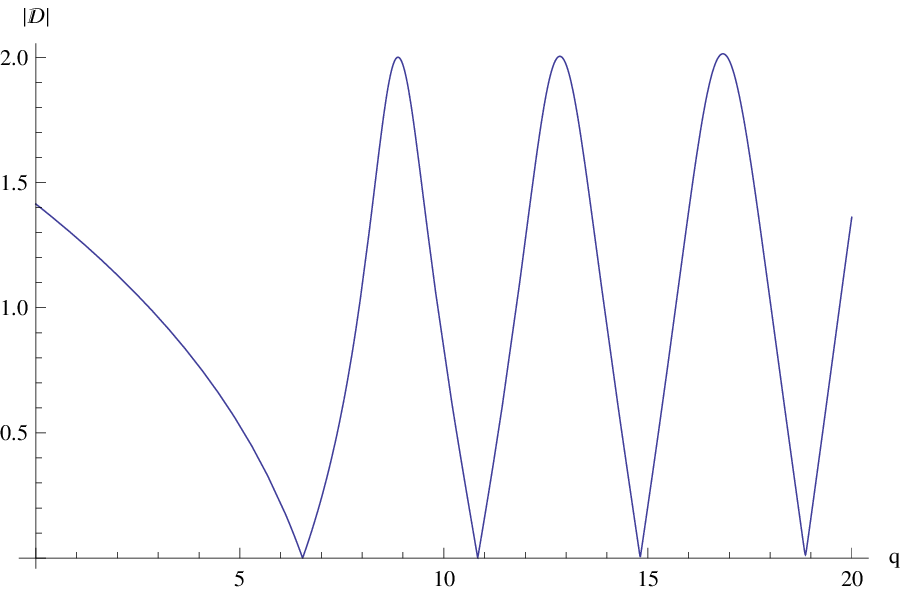}
\caption{The plots of $|\mathcal{D}(0,k)|$ as a function of $q$ for fixed k. The left plot for $k=1$; the right plot for $k=5$.  }
\label{fig12}\end{figure}

To study the behavior of $|\mathcal{D}|$ in the full $(k,q)$ plane, we then fix the momentum $k$ and plot $|\mathcal{D}|$ as a function of $q$. From fig.\ref{fig12}, we find that no Fermi surface exists for $k=1$ (the left plot). The first local minimum of $|\mathcal{D}|$ is above 0.18 which is too large to define a Fermi momenta, even in the approximate sense. When $k=5$, we find four Fermi surfaces which correspond to the four local minimums of $|\mathcal{D}|$ in the right plot. When the momentum increases further, we can find more and more Fermi surfaces and Fermi momenta.

\begin{figure}[tbp]
\begin{center}
\includegraphics[width=8.5cm]{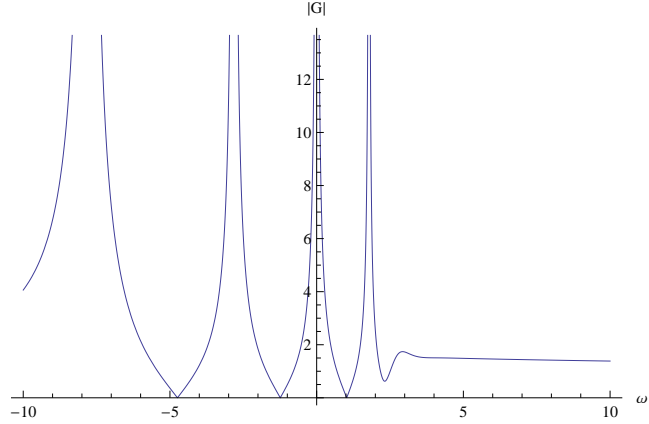}
\caption{The plots of $|G(\omega,k_F)|$ for $q=20$. The Fermi momentum is $k_F=9.42936683$. We see that multiple maxima spikes arise when $\omega$ varies. }
\end{center}
\label{fig13}\end{figure}

We have presented enough illustrative examples on the properties of the Green's function with vanishing $\omega$. It will also benefit us to study the Green's function for generic $\omega$. We choose $q=20$ and $k=9.42936683$ (the second Fermi momenta when $\omega=0$). In fig.13, we find multiple maxima spikes arise when $\omega$ varies from $-\infty$ to $\infty$. The spike emerging at $\omega=0$ corresponds to the Fermi surface with $k_F=9.42936683$ and the right spike corresponds to the Fermi surface with $k_F< 9.42936683$. When we increase k further, the left spikes will gradually cross the vertical axis and move to the right plane. By definition, the Fermi surfaces are defined by the poles of the Green's function with vanishing $\omega$. Thus, we will obtain one more Fermi surface when a spike in the left plane crosses the vertical axis. In short, we will obtain four Fermi surfaces for $q=20$ which is compatible with the information given in fig.\ref{fig10}.

\begin{figure}[tbp]
\includegraphics[width=5cm]{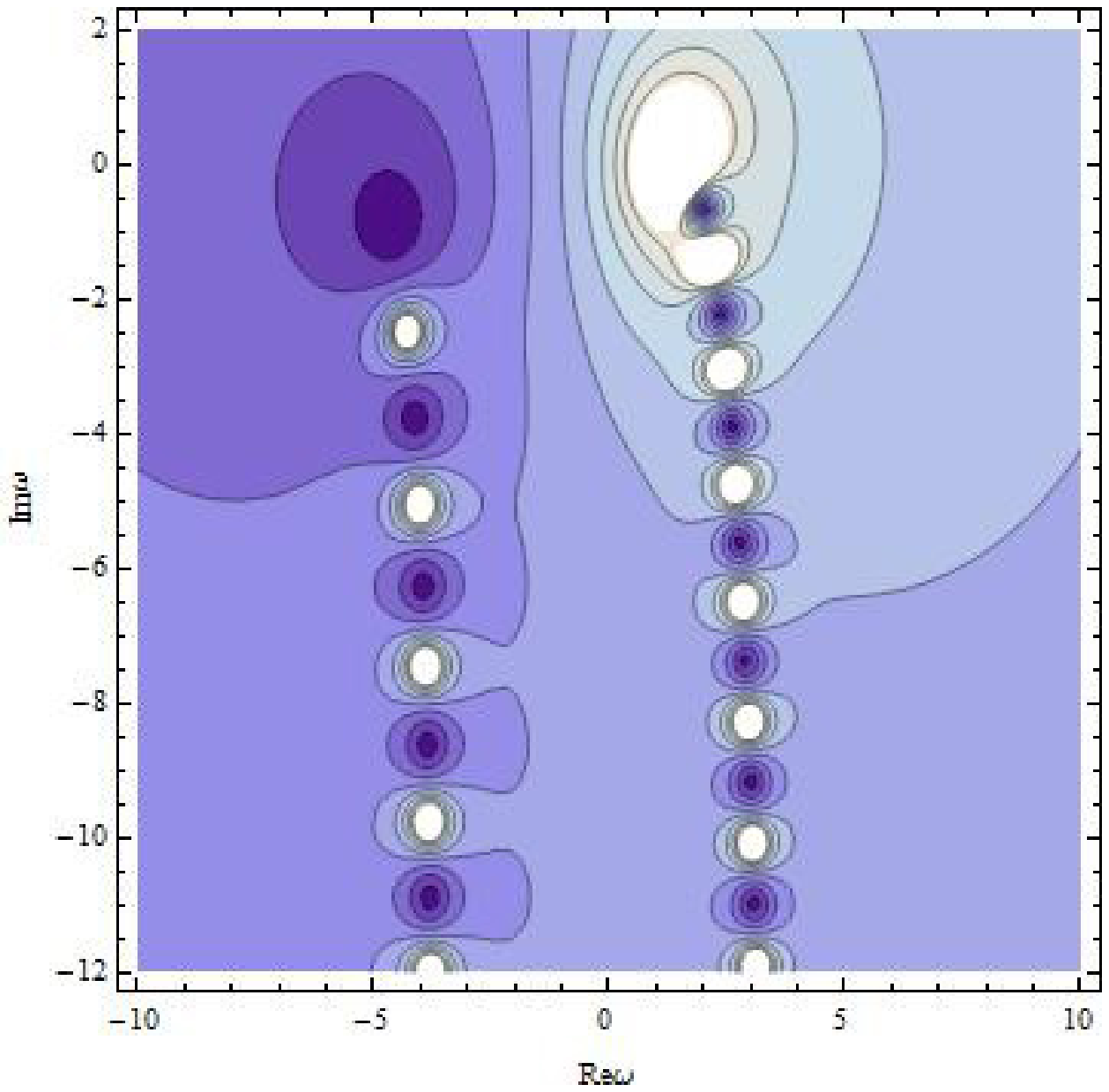}
\includegraphics[width=5cm]{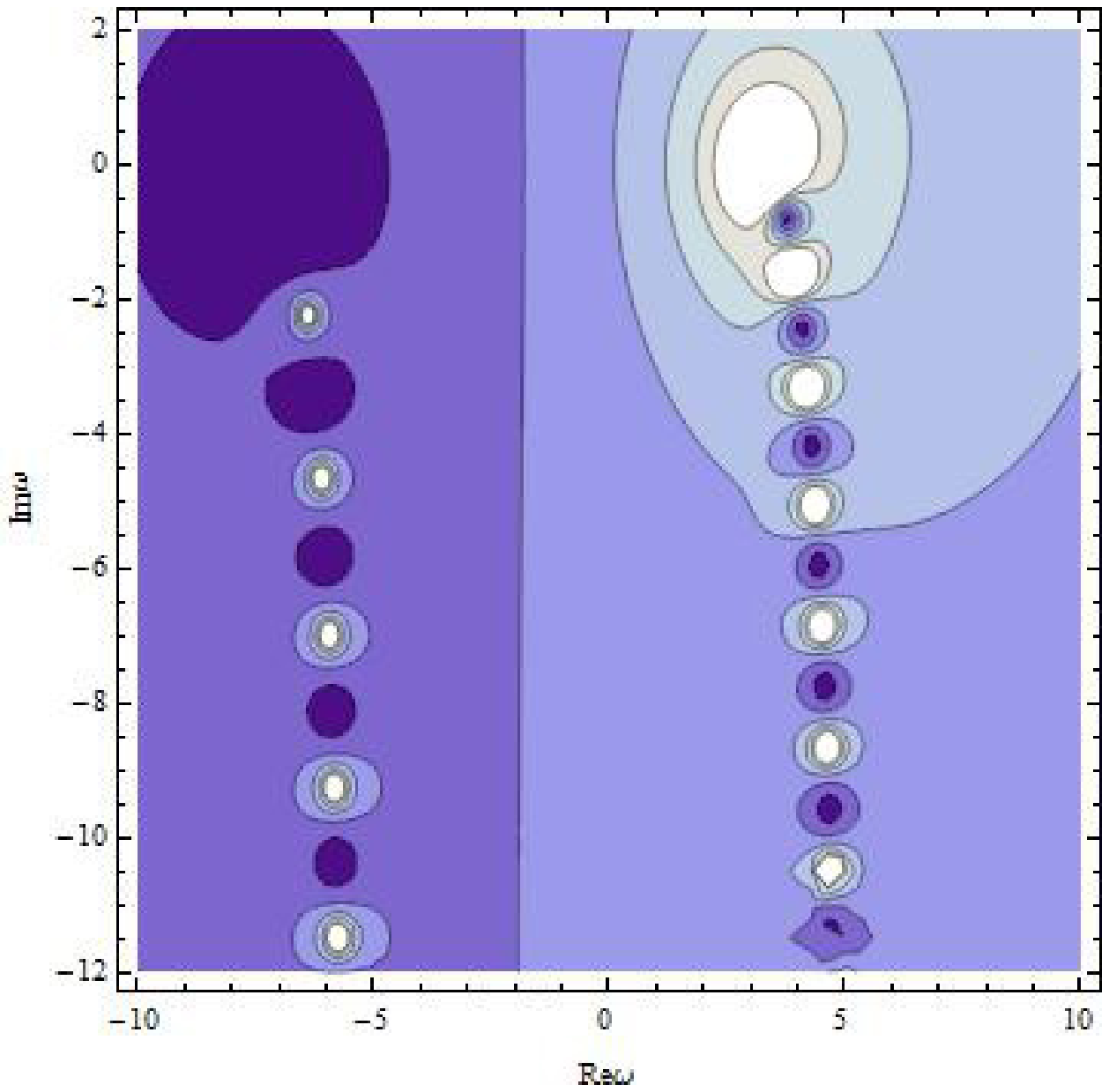}
\includegraphics[width=5cm]{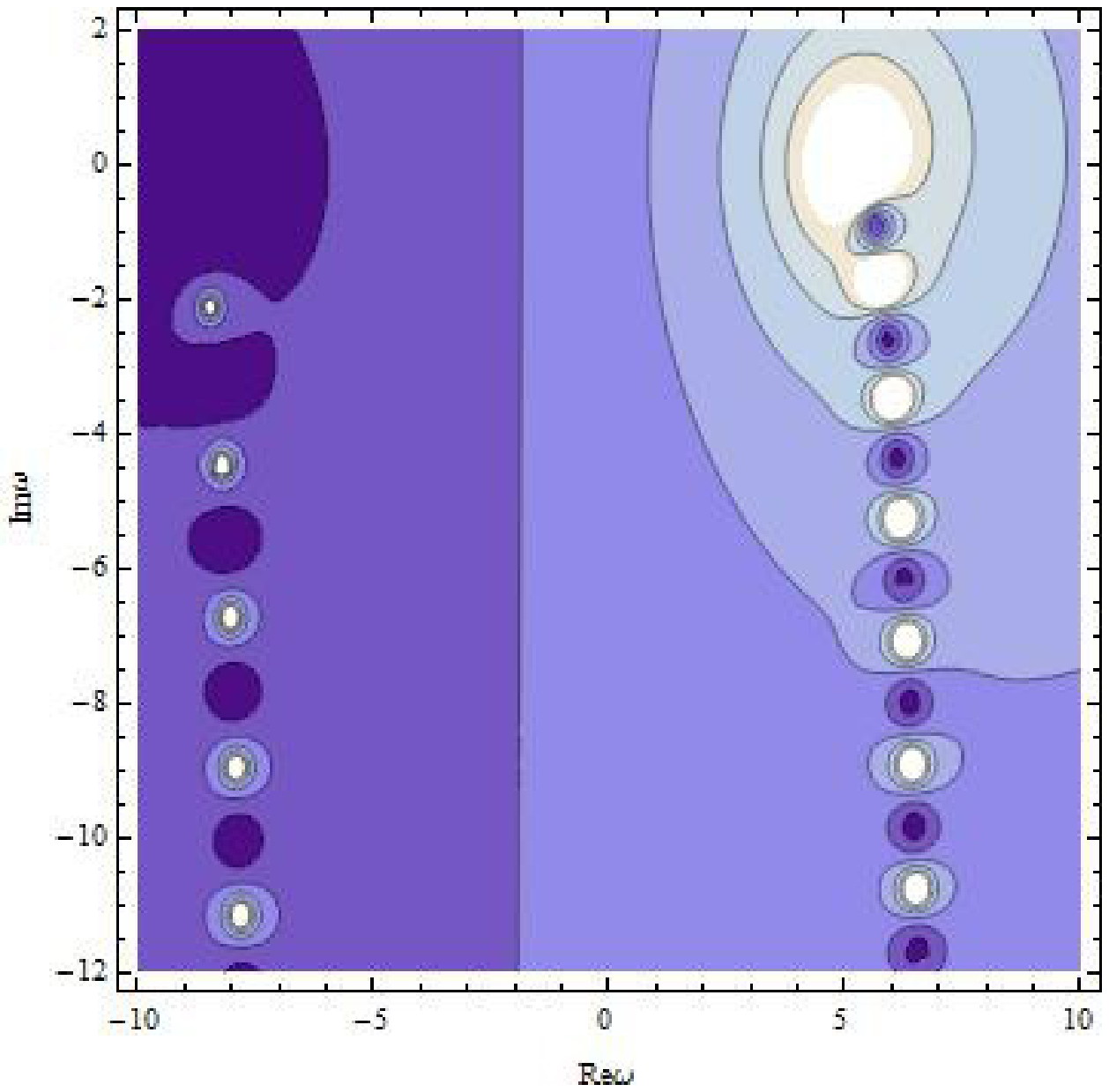}
\caption{The contour plots of $|G(\omega,k)|$ in the plane of complex $\omega$ for fixed k. The spinor charge is $q=2$. The left plot for $k=3$, the middle plot for $k=5$ and the right plot for $k=7$.  All the poles (white dots) and zeros (dark dots) occur in the lower half plane.}
\label{fig14}\end{figure}

In the end, we present contour plots for the Green's function $|G(\omega,k)|$ in the plane of complex $\omega$. In fig.\ref{fig14}, we fix $q=2$ and find that there are also two branches of poles and zeros. However, they are roughly parallel with the imaginary axis and never join together. From the left plot to the right plot, the momentum increases and the poles move towards both sides of the real axis in translational motion. All these features are crucially different from the $\delta=1$ case. Nevertheless, all the poles and zeros occur at the lower half plane, signifying that the dilatonic background is stable against fermionic perturbations as well as the $\delta=1$ case.

\subsection{Case 3: $\delta=3$}
When $\delta=3$, the equation of motion can be transformed into Heun's equation and the wave functions are solved in terms of Heun's functions:
\bea u_-(r)&=&(\tilde{r}-1)^{\alpha_1+\frac 12}(\tilde{r}+a^\ast)^{\alpha_2+\frac 12}(\tilde{r}+a)^{\alpha_3+\frac 12}H\ell(a,b_1;2,\beta+1,\gamma+1,\epsilon+1;z),\nn\\
 u_+(r)&=&ic(\tilde{r}-1)^{\alpha_1}(\tilde{r}+a^\ast)^{\alpha_2}(\tilde{r}+a)^{\alpha_3}H\ell(a,b_2;0,\beta,\gamma,\epsilon;z), \label{eq35}\eea
The notations are specified by
\be a=\frac{1-\sqrt{3}i}{2},\ b_1=\frac{3-\sqrt{3}i}{6}(3+\tilde{k}^2+2i\tilde{\mu}_q),\ b_2=\frac{3-\sqrt{3}i}{6}\tilde{k}^2,\ c=\frac{3i-\sqrt{3}}{2\tilde{k}}  \label{eq36}\ee
\be \alpha_1=-\frac{i\omega}{3},\quad \alpha_2=\frac{i-\sqrt{3}}{6}\tilde{\omega}-\frac{\sqrt{3}}{3}\tilde{\mu}_q,\quad \alpha_3=\frac{i+\sqrt{3}}{6}\tilde{\omega}+\frac{\sqrt{3}}{3}\tilde{\mu}_q,\quad \beta=\frac 12,\label{eq37}\ee
\be \gamma=\frac 12-\frac{2}{3}i\tilde{\omega},\qquad \epsilon=\frac 12+\frac{i-\sqrt{3}}{3}\tilde{\omega}-\frac{2\sqrt{3}}{3}\tilde{\mu}_q,\qquad z=\frac{3-\sqrt{3}i}{6}(1-\tilde{r}). \label{eq38}\ee
The Green's function can be read off from the asymptotical behavior of $u_\pm(r)$:
\be G(\omega,k)=i\frac{1-\mathcal{R}(\omega,k)}{1+\mathcal{R}(\omega,k)},\quad \mathcal{R}(\omega,k)=\frac{c\cdot H\ell(a,b_2;0,\beta,\gamma,\epsilon;\frac{3-\sqrt{3}i}{6})}{H\ell(a,b_1;2,\beta+1,\gamma+1,\epsilon+1;\frac{3-\sqrt{3}i}{6})}  \label{eq39}\ee
\begin{figure}[tbp]
\includegraphics[width=7cm]{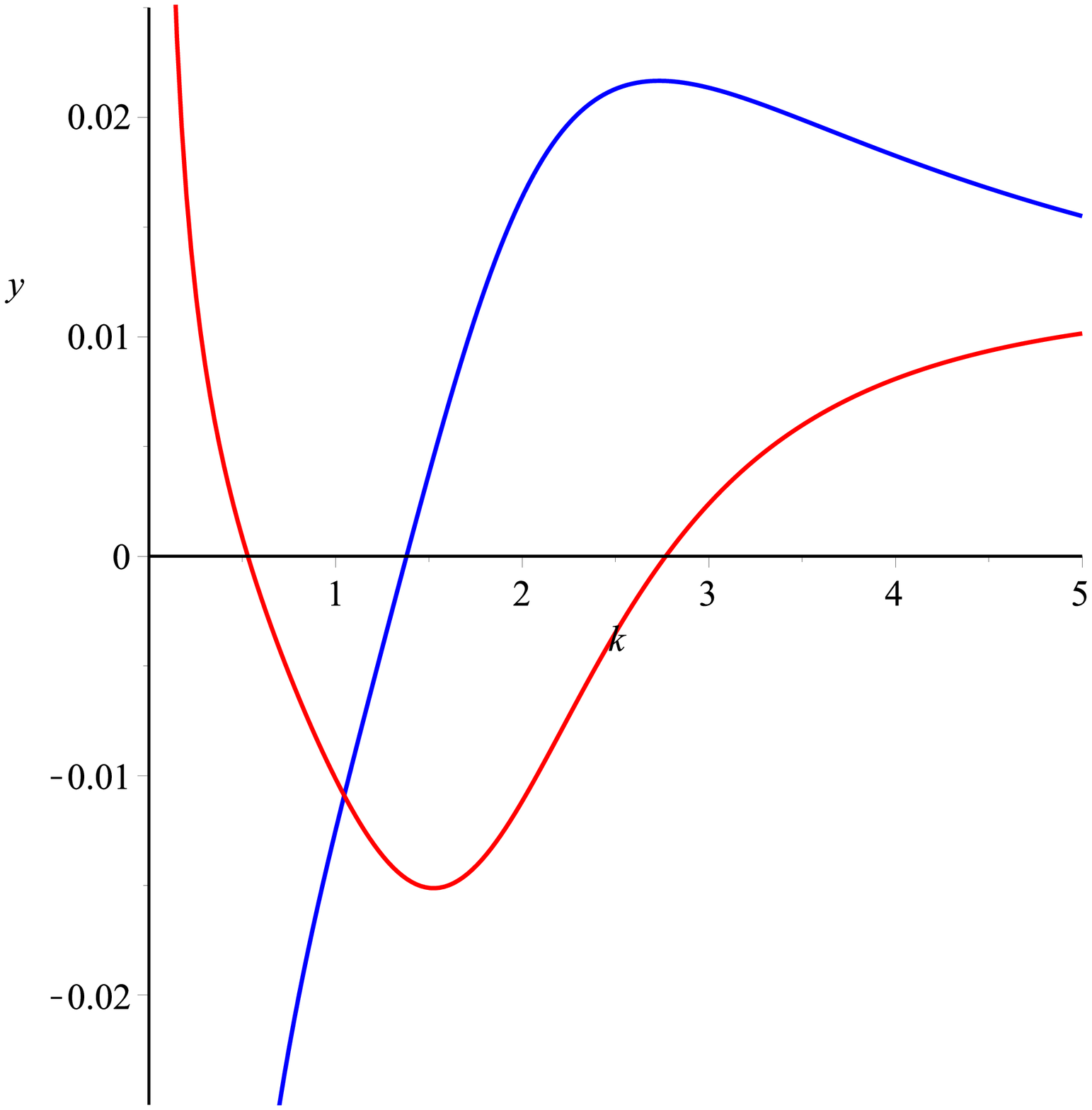}
\includegraphics[width=7cm]{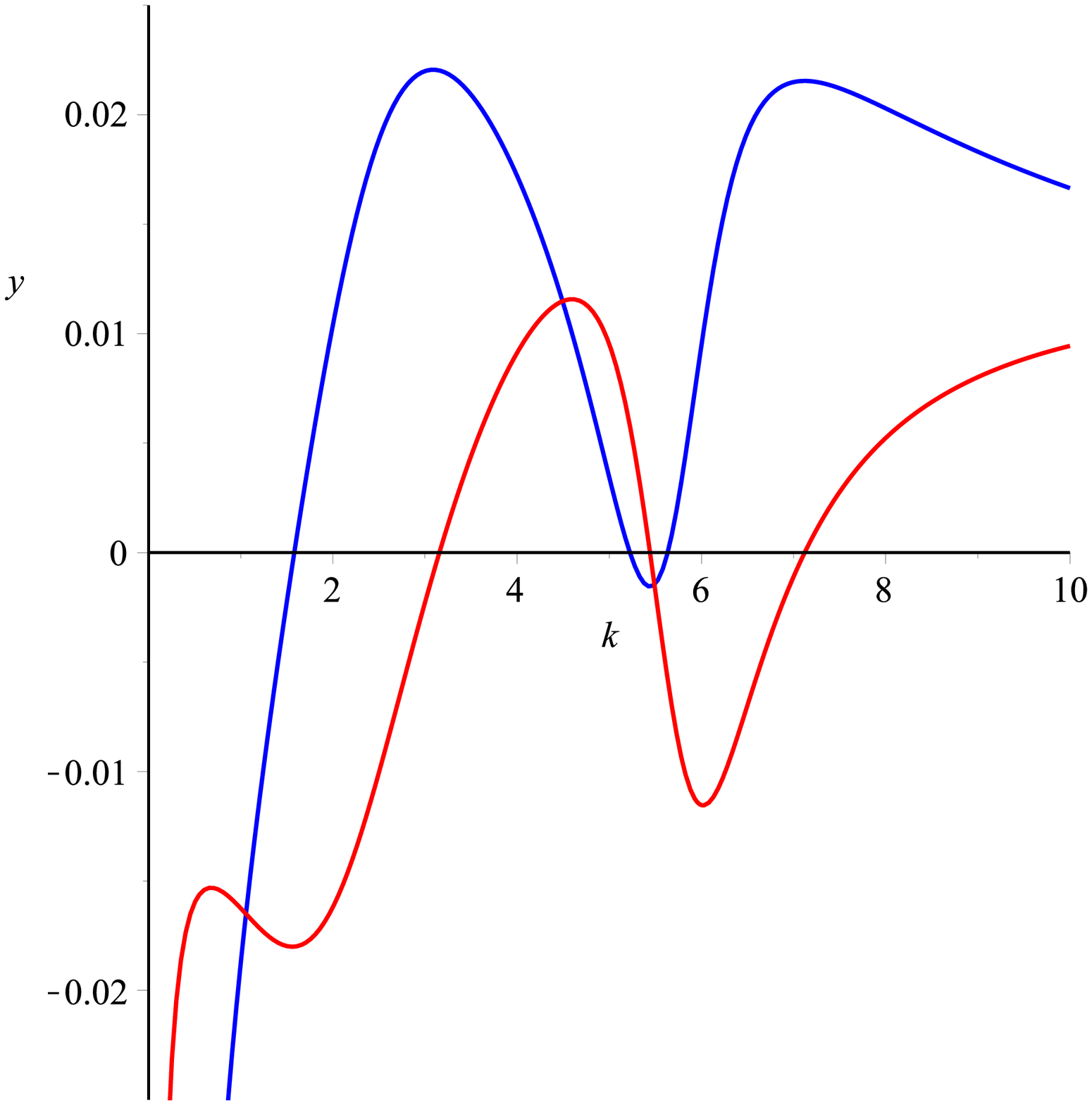}
\caption{These are the plots of the real (blue lines) and imaginary (red lines) part of $\mathcal{D}(0,k)$ as functions of k for fixed q. In the left plot, q=5 and no Fermi surfaces exist; in the right plot, q=10 and a Fermi surface emerges at $k_F=5.442223815$.  }
\label{fig15}\end{figure}
\begin{figure}[tbp]
\includegraphics[width=7cm]{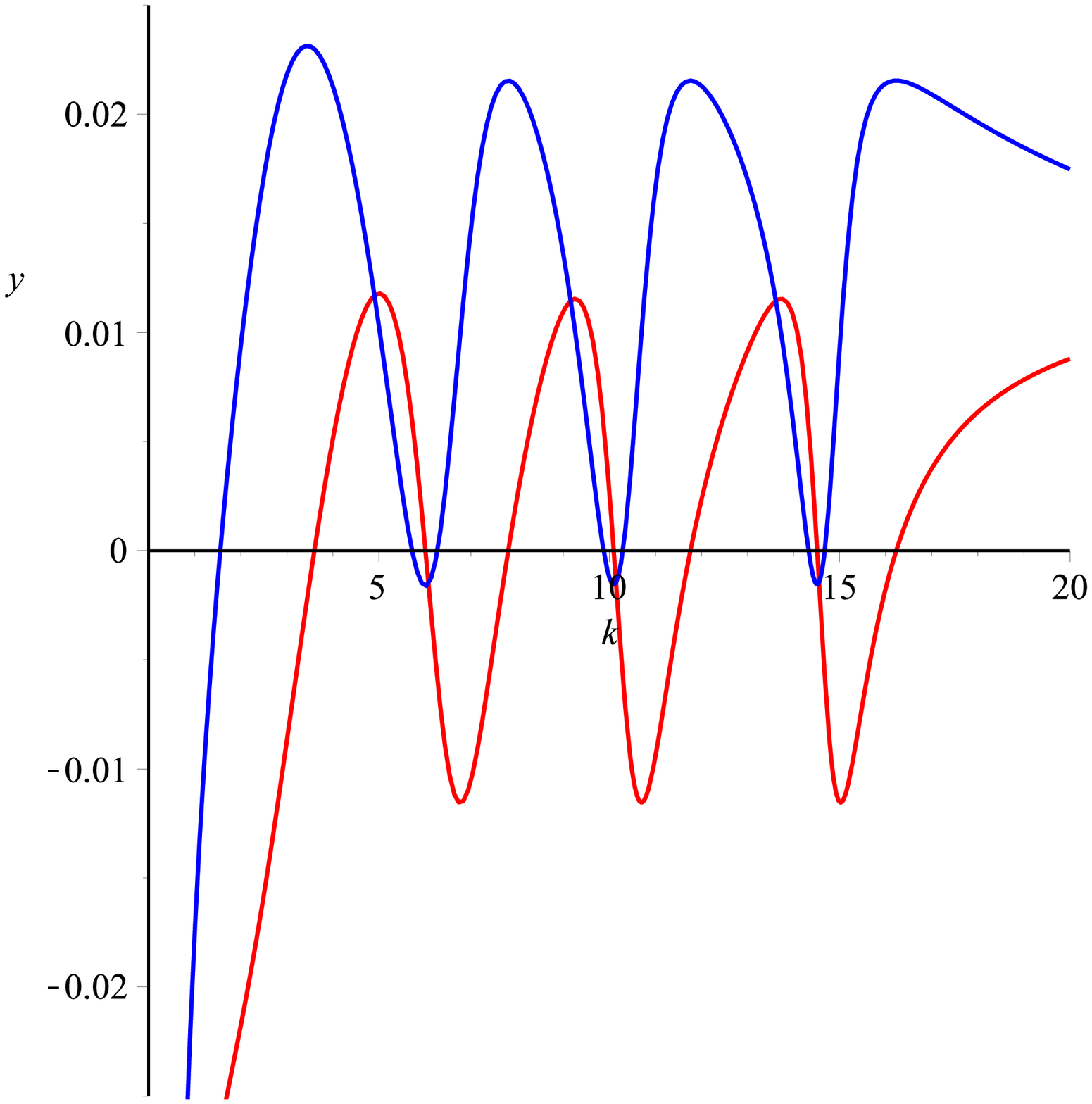}
\includegraphics[width=7cm]{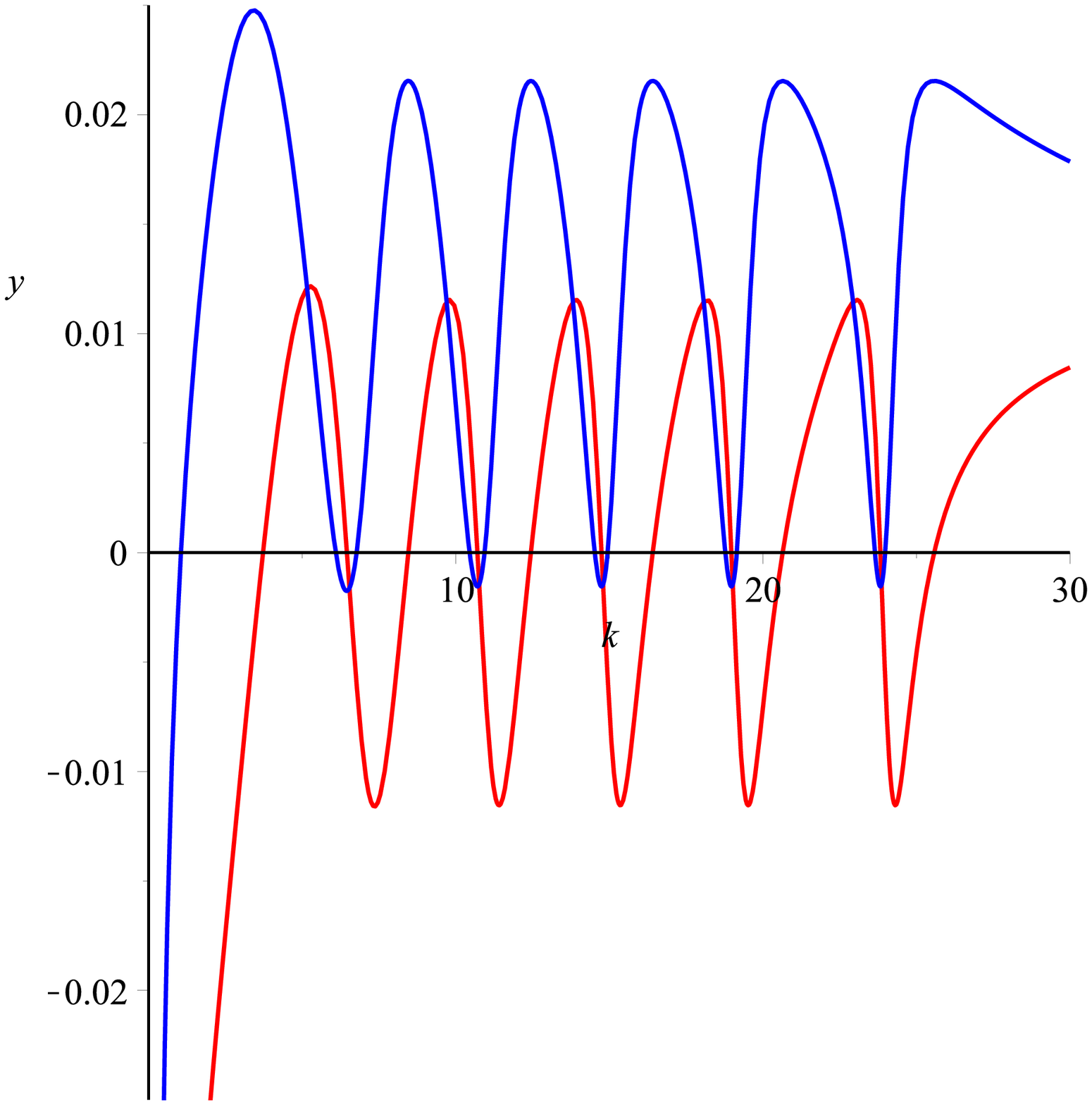}
\caption{These are the plots of the real (blue lines) and imaginary (red lines) part of $\mathcal{D}(0,k)$ as functions of k for fixed q. In the left plot, q=20 and three Fermi surfaces occur at $k_1=6.009538958,\ k_2=10.09652143$; in the right plot, q=30 and five Fermi surfaces occur at $k_1=6.456785816,\ k_2=10.70660293,\ k_3=14.74852806,\ k_4=18.97734513,\ k_5=23.82808183$.}
\label{fig16}\end{figure}

\begin{figure}[tbp]
\includegraphics[width=7cm]{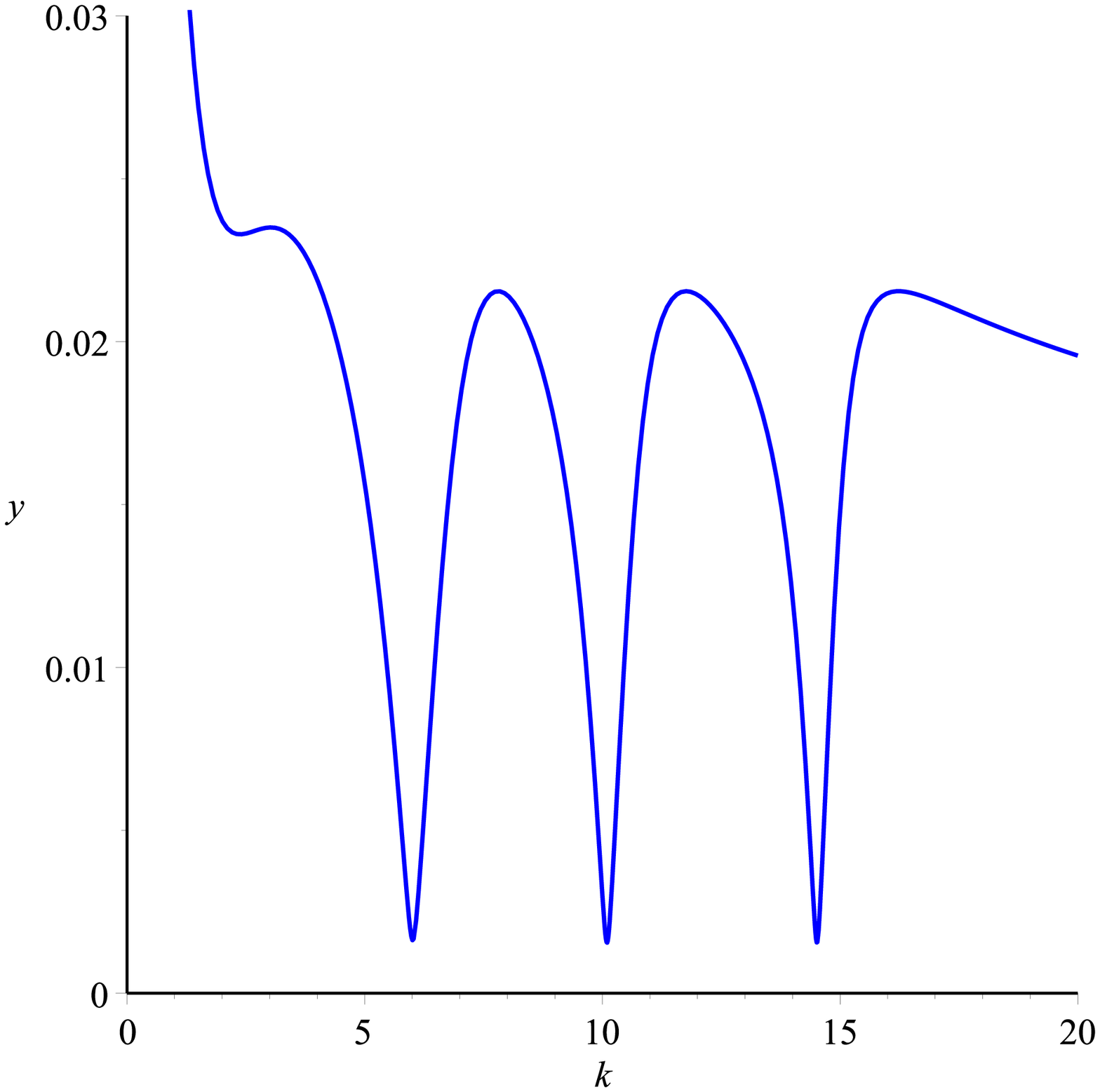}
\includegraphics[width=7cm]{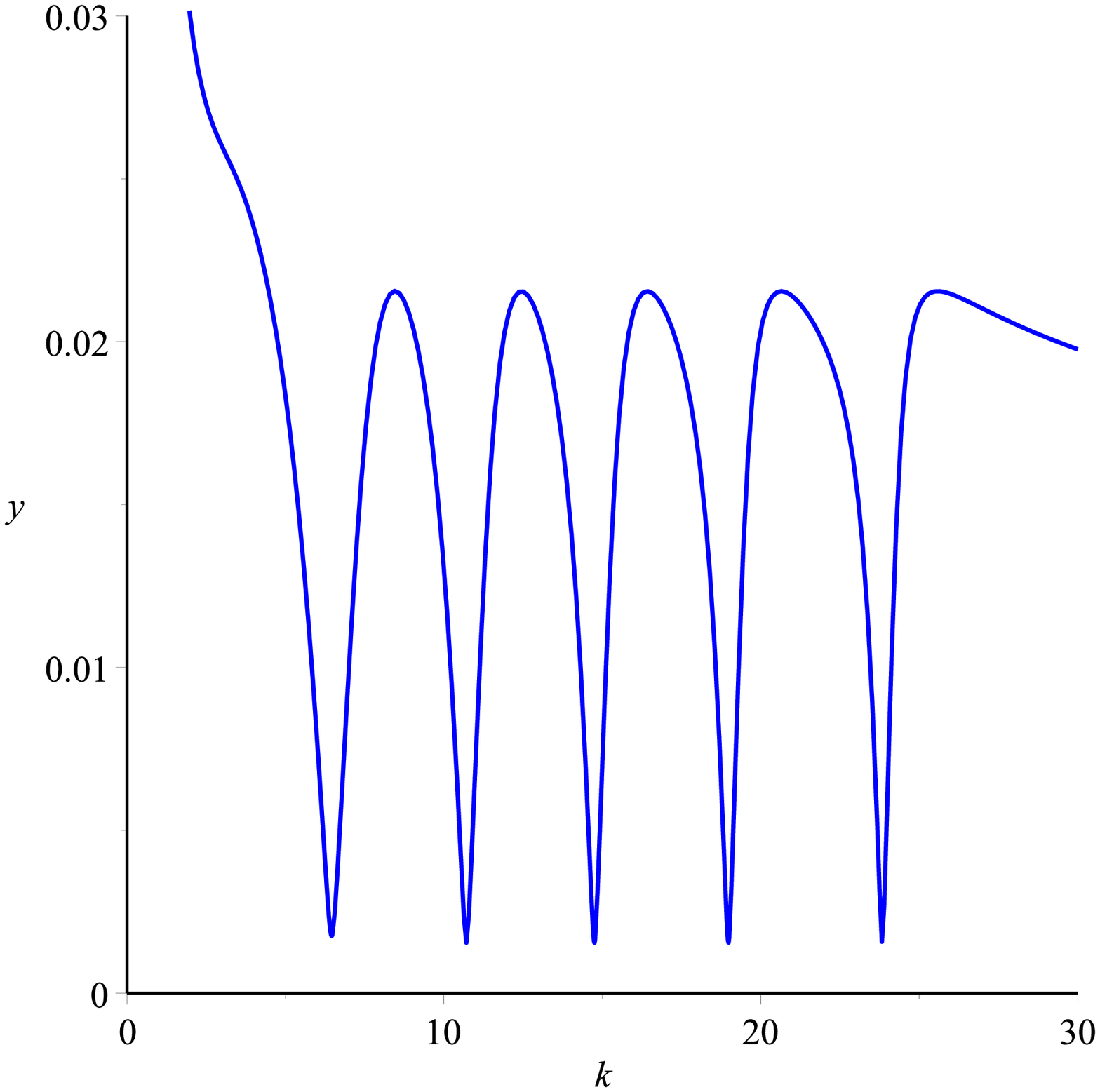}
\caption{The plots of $|\mathcal{D}(0,k)|$ for fixed q. The left plot for $q=20$; the right plot for $q=30$.  }
\label{fig17}\end{figure}

To search Fermi surfaces, we follow the strategy outlined in above two subsections and investigate the denominator of the Green's function with vanishing $\omega$. In fig.\ref{fig15}, we present two graphs in which the absolute value $|\mathcal{D}(0,k)|$ is ploted as functions of k. In the left plot, we set $q=5$ but find no Fermi surfaces. In the right plot, we set $q=10$ and find one Fermi surface at $k_F=5.442223815$. Notice that at this Fermi momenta, the real part of the denominator $Re(\mathcal{D})\sim -1.554\times 10^{-4}$ which does not literally vanish. Thus, the Fermi surface we find is only defined approximately. By increasing the spinor charge $q$, we find more Fermi surfaces and Fermi momenta, see fig.\ref{fig16} for the illustrative examples. The generic feature we observe is that all the Fermi surfaces arising in fig.\ref{fig16} can only be defined by the local minimums of the denominator $|\mathcal{D}|$. Actually, in all these plots, we have $|\mathcal{D}|\sim 0.0015-0.0017$ which is sufficient to define Fermi surfaces for practical purpose, see also fig.\ref{fig17}.

\begin{figure}[tbp]
\includegraphics[width=7cm]{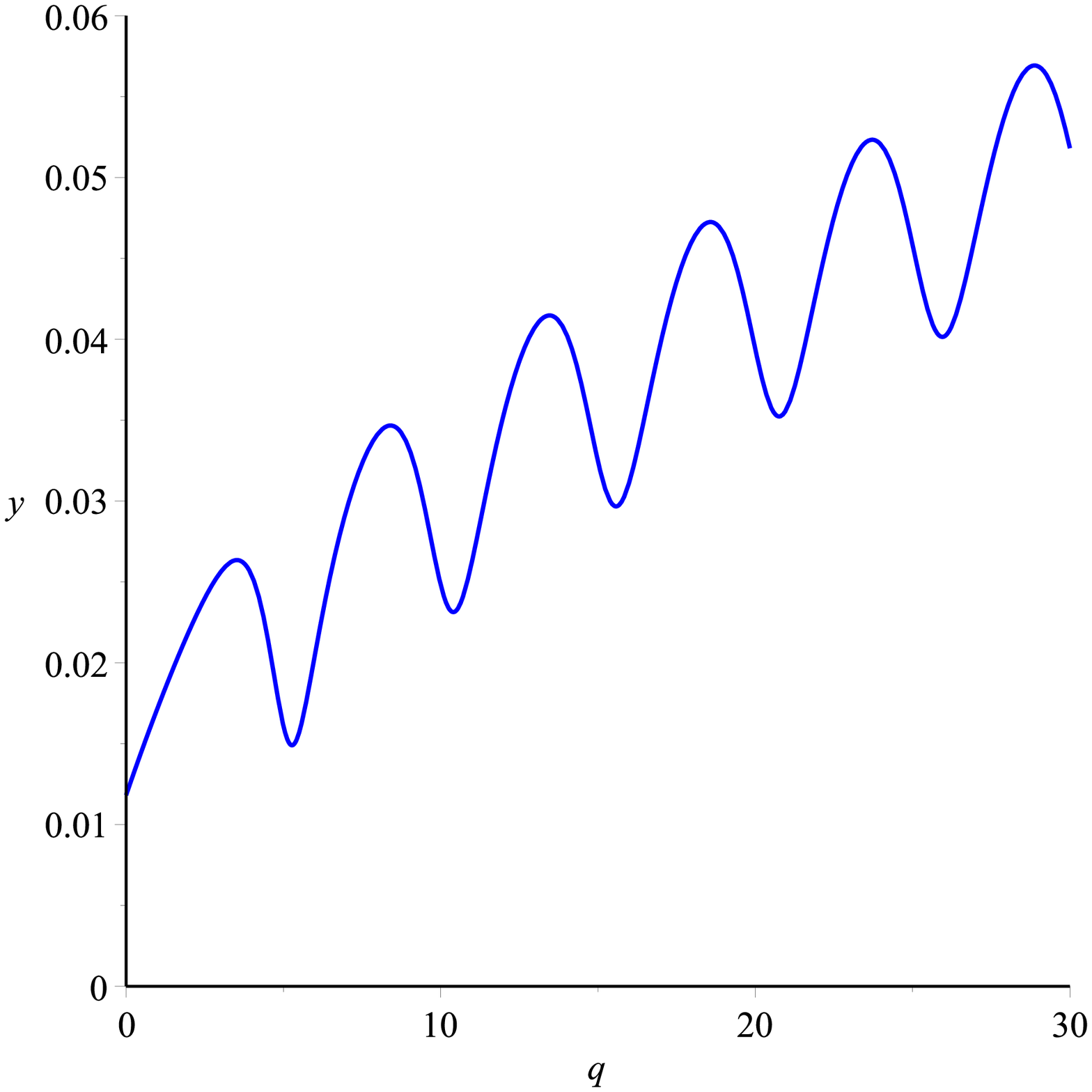}
\includegraphics[width=7cm]{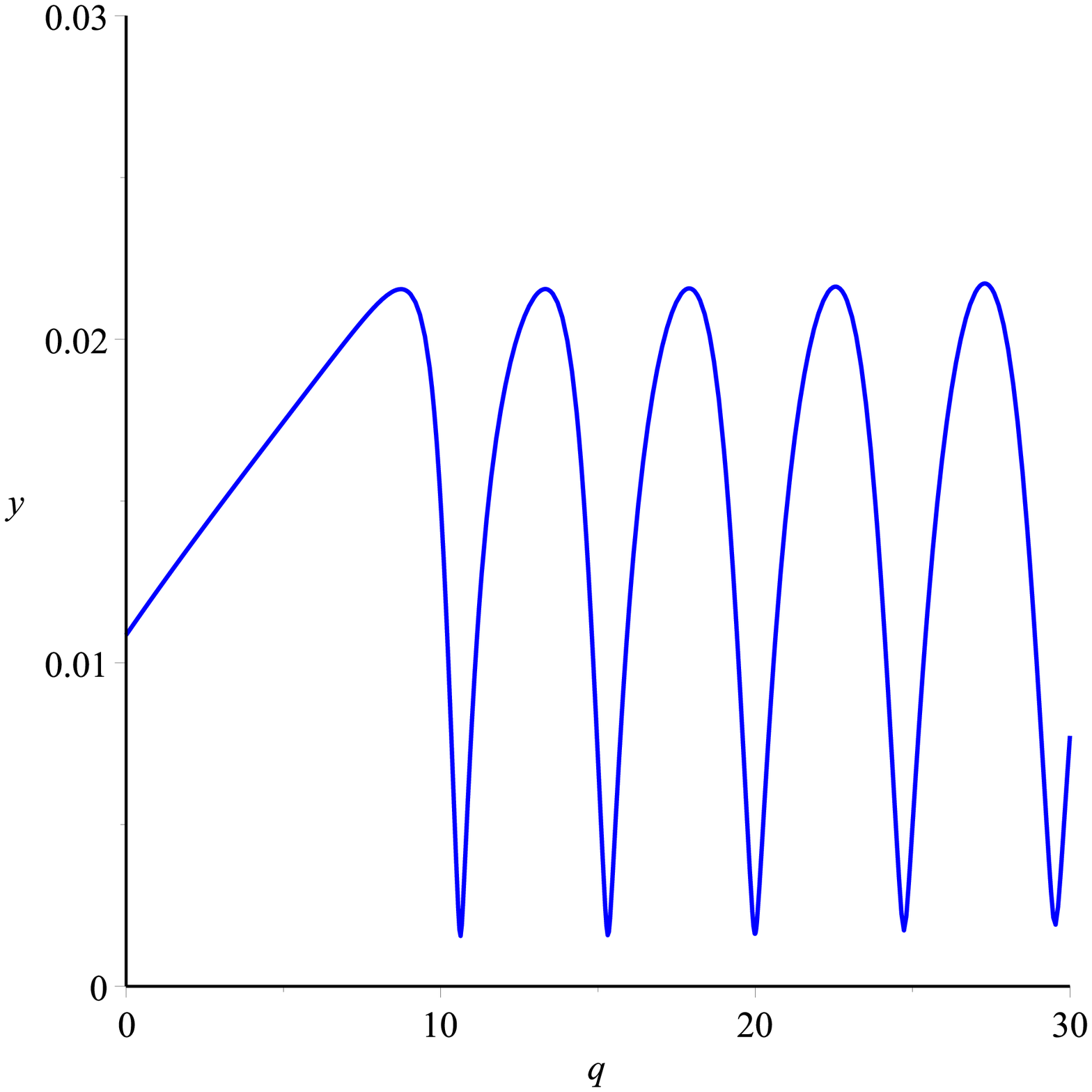}
\caption{The plots of $|\mathcal{D}(0,k)|$ as a function of $q$ for fixed k. The left plot for $k=1$; the right plot for $k=6$.  }
\label{fig18}\end{figure}

In fig.\ref{fig18}, we draw $|\mathcal{D}|$ for fixed k while the charge $q$ varies from 0 to $\infty$. When $k$ is small, for example $k=1$, there are no Fermi surfaces. However, when $k$ increases and becomes larger, more and more Fermi surfaces begin to emerge. For example, in the right plot, we have $k=6$ and find five Fermi surfaces. If the momentum $k$ increases further, the number of Fermi surface will be more than five.


In the end, we present a 3D plot for the Green's function $|G(0,k)|$ in the plane of $(k,q)$. It is easy to see that many sharp peaks arise in the lower half region characterized by $k\leq q$. The sharp peaks are associated with the effective Fermi surfaces and Fermi momenta. The fascinating rich structure of $|G|$ shows that a large number of Fermi surface emerge when the momenta $k$ and the spinor charge $q$ becomes sufficient large.

\section{Conclusions}
In this paper, we first construct charged dilatonic black holes in four dimensional anti-de-sitter space-time using Einstein-Maxwell-Dilaton theories. We require the thermal factor of the black holes has the form of:$\ h(r)=1-(r/r_h)^{\delta}$, with $\delta>0$. Hence, in the extremal limit the space-time becomes deformed AdS (with deformed factor given by $\Omega^2$) and the entropy of the black holes is vanishing. This is a fantastic feature which provides a more realistic holographic model for condensed-matter physics.
\begin{figure}[tbp]
\begin{center}
\includegraphics[width=8.5cm]{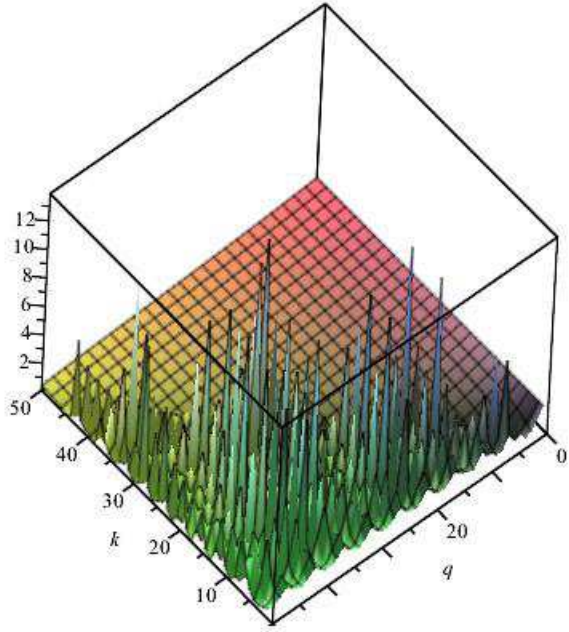}
\caption{The 3D plot of $|G(0,k)|$ in the plane of $(k,q)$. We find that multiple sharp peaks arise in the lower half region $k\leq q$. }
\end{center}
\label{fig19}\end{figure}

The most interesting result we report is that the Dirac equation of massless spinors can be exactly solved in those dilatonic black holes having $\delta=1,2,3,4$. Depending on the certain value of $\delta$, the wave functions can be solved analytically in terms of Whittaker functions ($\delta=1$), hypergeometric functions ($\delta=2$) or Heun's functions ($\delta=3,4$). According to holographic dictionaries, the retarded Green's functions can be read off
from the asymptotical behavior of the wave functions by imposing in-going boundary condition at the horizon. It is worth remarking that in our system, the Green's functions are exactly solved for generic frequency $\omega$ and momentum $k$ at finite temperature.

By investigating the Green's function with vanishing $\omega$, we find a fascinating rich structure of maxima spikes in the plane of $k$ and the spinor charge $q$, which strongly implies that many Fermi surfaces emerge in the charged dilatonic black holes. We present many examples to illustrate the new distinguishing properties found in these black holes. Our analytical results may provide a more elegant approach to study condensed-matter physics in the strongly coupled region using gauge/gravity duality.

\section*{Acknowledgement}

We are grateful to Si-Jie Gao and H. L\"{u} for useful discussions. Zhong-Ying Fan is supported by NSFC Grants NO.10975016, NO.11235003 and NCET-12-0054.

\section{Appendix\quad Case 4: $\delta=4$}
We also find that the Dirac equation is exactly solvable when $\delta=4$. The wave functions are solved in terms of Heun's functions:
\bea u_-(r)&=&(\frac{\tilde{r}-1}{\tilde{r}+i})^{\alpha_1+\frac 12}(\frac{\tilde{r}-i}{\tilde{r}+i})^{\alpha_2+\frac 12}(\frac{\tilde{r}+1}{\tilde{r}+i})^{\alpha_3+\frac 12}H\ell(2,b_1,2,\beta+1,\gamma+1,\epsilon+1;z),\nn\\ u_+(r)&=&\frac{\tilde{k}}{2}(\frac{\tilde{r}-1}{\tilde{r}+i})^{\alpha_1}(\frac{\tilde{r}-i}{\tilde{r}+i})^{\alpha_2}(\frac{\tilde{r}+1}{\tilde{r}+i})^{\alpha_3}
H\ell(2,b_2,0,\beta,\gamma,\epsilon;z),\label{eq42}\eea
where the notations are specified by
\be \alpha_1=-\frac{i\tilde{\omega}}{4},\quad\qquad \alpha_2=\frac 14(-i\tilde{\mu}_q+\tilde{\mu}_q+\tilde{\omega}),\quad\qquad \alpha_3=\frac{i\tilde{\omega}}{4}+\frac{i\tilde{\mu}_q}{2}. \label{eq41}\ee
\be \beta=\frac 12(1+i\tilde{\mu}_q+\tilde{\mu}_q+\tilde{\omega}),\quad \gamma=\frac{1-i\tilde{\omega}}{2},\quad \epsilon=\frac 12(1-i\tilde{\mu}_q+\tilde{\mu}_q+\tilde{\omega}),\label{eq41}\ee
\be b_1=3-\frac{i\tilde{k}^2}{2}+\tilde{\mu}_q+(1-i)\tilde{\omega},\qquad b_2=-\frac{i\tilde{k}^2}{2},\qquad z=(1-i)\frac{\tilde{r}-1}{\tilde{r}+i}, \label{eq43}\ee
The Green's function can be read off by
\be G(\omega,k)=i\frac{1-\mathcal{R}(\omega,k)}{1+\mathcal{R}(\omega,k)},\quad \mathcal{R}(\omega,k)=\frac{i\tilde{k}}{2}\frac{H\ell(2,b_2,0,\beta,\gamma,\epsilon;1+i)}{H\ell(2,b_1,\beta+1,\gamma+1,\epsilon+1;1+i)}. \label{eq44}\ee
Unfortunately, we have not found a mathematical package which can plot above Green's functions. Compared to hypergeometric functions, the Heun's functions are much less studied. In fact, they are not encoded in Mathematic package while in Maple package, only vert limited properties are encoded. Nevertheless, we obtain the Green's function exactly. The existence of Fermi surfaces can be ensured in numerics.


\begin{thebibliography}{99}

\bibitem{1}T. Faulkner, N. Iqbal, H. Liu, J. McGreevy, D. Vegh, {\it Strange metal transport realized by gauge/gravity duality}, Science {\bf 329} 1043 (2010).

\bibitem{2}H. Liu, J. McGreevy, D. Vegh, {\it Non-Fermi liquids from holography}, Phys.\ Rev.\ D {\bf 83}, 065029 (2011).

\bibitem{3}T.Faulkner, H.Liu, J.McGreevy, and D.Vegh, {\it Emergent quantum criticality, Fermi surfaces and $\mbox{AdS}_2$}, Phys.\ Rev.\ D {\bf 83},125002 (2011).

\bibitem{4}T. Hartman and S. A. Hartnoll, {\it Cooper pairing near charged black holes}, JHEP {\bf 1006}, 005 (2010).

\bibitem{5}S.S. Gubser and J. Ren, {\it Analytic fermionic Green's functions from holography}, Phys.\ Rev.\ D {\bf 86}, 046004 (2012).

\bibitem{6}H.L\"{u}, Zhao-Long Wang, {\it Exact Green's Function and Fermi Surfaces from Conformal Gravity}, Phys.\ Lett.\ B {\bf 718} (2013) 1536-1542.

\bibitem{7}Jun Li, Hai-Shan Liu, H.L\"{u} and Zhao-Long Wang, {\it Fermi Surfaces and Analytic Green's Functions from Conformal Gravity}, JHEP {\bf 02} (2013) 109.

\bibitem{8}N. Iqbal, H. Liu, and M. Mezei, {\it Semi-local quantum liquids}, JHEP {\bf 1204}, 086 (2012).

\bibitem{9}S.A. Hartnoll, J. Polchinski, E. Silverstein and D. Tong, {\it Towards strange metallic holography }, JHEP {\bf 1004}, 120 (2010).

\bibitem{10}S.A. Hartnoll and A. Tavanfar, {\it Electron stars for holographic metallic criticality}, Phys.\ Rev.\ D {\bf 83},046003 (2011).

\bibitem{11}S.A. Hartnoll and P. Petrov, {\it Electron star birth: A continuous phase transition at nonzero density}, Phys.\ Rev.\ Lett. {\bf 106}, 121601 (2011).

\bibitem{12}S.A. Hartnoll, D.M. Hofman, D.Vegh, {\it Stellar spectroscopy: Fermions and holographic Lifshitz criticality}, JHEP {\bf 1108}, 096 (2011).

\bibitem{13}Danning. Li, Mei. Huang, {\it Dynamical holographic QCD model for glueball and light meson spectra}, JHEP {\bf 11} (2013) 88.

\bibitem{14}Rong-Gen Cai, S. Chakrabortty, Song. He and Li. Li, {\it Some aspects of QGP phase in a hQCD model}, JHEP {\bf 02} (2013) 068.





\end{thebibliography}
\end{document}